\documentclass[floats,floatfix,amssymb,prd,twocolumn,superscriptaddress,nofootinbib,preprintnumbers]{revtex4-2}
\usepackage{graphicx,amssymb,amsmath,amsthm,amsfonts,epsfig,epsf}
\usepackage[citecolor=blue,colorlinks=true]{hyperref}
\usepackage{breakurl}
\usepackage[usenames]{color}
\usepackage{bm}
\usepackage{multirow}
\usepackage{gensymb}
\usepackage{booktabs}
\usepackage{verbatim}
\usepackage{ORCIDinREVTeX}
\newcommand{\milan}{\affiliation{Dipartimento di Fisica ``G. Occhialini'', 
Universit\'a degli Studi di Milano-Bicocca, Piazza della Scienza 3, 20126 Milano, Italy}}
\newcommand{\infnmilan}{\affiliation{INFN, Sezione di Milano-Bicocca, 
Piazza della Scienza 3, 20126 Milano, Italy}}

\begin{document}

	\title{Taming systematics in distance and inclination measurements \\ with gravitational waves: role of the detector network and higher-order modes
 }

\author{Adriano Frattale Mascioli}
\email{Corresponding author, email:\,adriano.frattalemascioli@uniroma1.it}
\orcid{0000-0002-0155-3833}
\affiliation{Dipartimento di Fisica, ``Sapienza'' Universit\`a di Roma, Piazzale Aldo Moro 5,  00185, Roma, Italy}
\affiliation{Dipartimento di Fisica, Sezione INFN Roma, Piazzale Aldo Moro 5,  00185, Roma, Italy}

\author{Francesco Crescimbeni}
\orcid{0009-0001-4088-5443}
\affiliation{Dipartimento di Fisica, ``Sapienza'' Universit\`a di Roma, Piazzale Aldo Moro 5,  00185, Roma, Italy}
\affiliation{Dipartimento di Fisica, Sezione INFN Roma, Piazzale Aldo Moro 5,  00185, Roma, Italy}
      
\author{Costantino Pacilio}
\orcid{0000-0002-8140-4992}
\milan
\infnmilan

\author{Paolo Pani}
\orcid{0000-0003-4443-1761}
\affiliation{Dipartimento di Fisica, ``Sapienza'' Universit\`a di Roma, Piazzale Aldo Moro 5,  00185, Roma, Italy}
\affiliation{Dipartimento di Fisica, Sezione INFN Roma, Piazzale Aldo Moro 5,  00185, Roma, Italy}

\author{Francesco Pannarale}
\orcid{0000-0002-7537-3210}
\affiliation{Dipartimento di Fisica, ``Sapienza'' Universit\`a di Roma, Piazzale Aldo Moro 5,  00185, Roma, Italy}
\affiliation{Dipartimento di Fisica, Sezione INFN Roma, Piazzale Aldo Moro 5,  00185, Roma, Italy}
	
\begin{abstract}
Gravitational-wave (GW) observations of compact binaries have the potential to unlock several remarkable applications in astrophysics, cosmology, and nuclear physics through accurate measurements of the source luminosity distance and inclination. However, these parameters are strongly correlated when performing parameter estimation, which may hamper the enormous potential of GW astronomy. We comprehensively explore this problem by performing Bayesian inference on synthetic data for a network of current and planned second-generation GW detectors, and for the third-generation interferometer Einstein Telescope~(ET). We quantify the role of the network alignment factor, detector sensitivity, and waveform higher-order modes in breaking this degeneracy. We discuss the crucial role of the binary mass ratio: in particular, we find that ET can efficiently remove the error in the distance as long as the compact binary is asymmetric in mass.
	\end{abstract}

 \preprint{ET-0079A-25; DCC:P2500127; VIR-0347A-25}
	
	\maketitle

\tableofcontents

	\section{Introduction}\label{sec:intro}
The discovery of gravitational waves~(GWs) by the LIGO-Virgo Collaboration~\cite{LIGOScientific:2016aoc,LIGOScientific:2016vbw} marked a historical moment in science, and inaugurated a series of remarkable results.
The first detection of a binary neutron star~(BNS) merger, GW170817, constrained the behavior of nuclear matter to an unprecedented level, and the multimessenger observation of a coincident electromagnetic signal shed light on the origin of short gamma-ray bursts~\cite{LIGOScientific:2017vwq, LIGOScientific:2017ync, LIGOScientific:2017zic} and on heavy-element nucleosynthesis~\cite{GW170817_nucl}.
In parallel, the approximately 90 compact binary coalescences detected so far have provided unique information about the properties of binary black hole~(BBH) populations and their astrophysical environment~\cite{LIGOScientific:2020zkf} that is essential for elucidating binary formation scenarios. Further, they allowed for novel tests of gravity in the highly dynamical, strong-field regime, confirming Einstein's General Relativity and severely constraining various alternative theories of gravity~\cite{test_GR_1, test_GR_2, test_GR_3}.

Despite the milestones achieved in the first three observing runs, GW astronomy is still in its early stages. The fourth observing run~(O4) started in May 2023, and the fifth observing run (O5) is planned for late 2027,\footnote{\url{https://observing.docs.ligo.org/plan/}.} with further enhanced sensitivities for the LIGO~\cite{LIGOScientific:2014pky, LIGO-Curves}, Virgo~\cite{VIRGO:2014yos}, and KAGRA~\cite{KAGRA:2018plz}
detectors.  In the 2030s, a fifth detectors, LIGO-India~\cite{Saleem:2021iwi}, will be joining the network.
While the horizon of these detectors is limited to the local universe, planned third-generation~(3G) detectors such as the Einstein Telescope (ET)~\cite{Hild:2010id,Punturo:2010zz,Abac:2025saz} and Cosmic Explorer~\cite{CE1,CE2} will allow for high-precision detections and will reveal GW signals from sources at distances well beyond the reach of current detectors~\cite{3G-science-case}.
In parallel, the ESA-NASA space mission LISA will open the mHz GW window populated by supermassive BBHs across the cosmic history~\cite{LISA:2017pwj,LISA:2024hlh}, while recent PTA data measurements showed the first evidence for the detection of a stochastic background of GWs at the nHz frequencies~\cite{PTA1, PTA2}.

The detection and parameter estimation of compact binary coalescence (CBC) events relies on waveform models of the signals, to filter the data, and infer intrinsic and extrinsic properties of the source~\cite{LVC_guide, Chatziioannou:2024hju}. 
Two extrinsic parameters that describe the CBC GW signal observed by a detector are the luminosity distance $d_L$ and the inclination angle $\iota$ between the orbital angular momentum and the line of sight.\footnote{In this paper, we will not address precession, where a more convenient angle to use is the one between the total angular momentum and the line of sight, $\theta_{\rm JN}$.  In the absence of precession, $\iota\equiv \theta_{\rm JN}$.}

An accurate measurement of these parameters is crucial in applications ranging from cosmology to nuclear physics.
For example, a combination of the luminosity distance measurement from a GW signal and the redshift measurement from an electromagnetic counterpart, from the same event, can be used to infer the value of the Hubble constant~\cite{Holz:2005df, Dalal:2006qt}, as done for GW170817~\cite{LIGOScientific:2017adf}.
More in general, future GW observations can play a crucial role in testing the current $\Lambda$CDM cosmological model and in searching for new physics, because of their powerful property of being directly sensitive to the luminosity distance.
This implies the possibility of measuring the Hubble constant with a statistical approach even when an electromagnetic counterpart is not present~\cite{Schutz:1986gp, LIGOScientific:2018gmd, LIGOScientific:2019zcs}.
Distance measurements are also pivotal in constraining the equation of state of neutron stars~\cite{Chatziioannou:2020pqz, Gupta:2022qgg}.
Indeed, while GW detections of BNSs allow for measurements of the neutron star tidal deformability~\cite{Yagi:2013bca, Cardoso:2017cfl, Dietrich:2020efo} and the masses in the detector frame, in order to recover the equation of state one needs information about the masses in the source frame, which inevitably requires knowledge of the source redshift. The latter cannot be directly extracted from the waveform. However, if an accurate measurement of the distance is available, the redshift can be inferred assuming a cosmological model~\cite{Planck:2018vyg} or, for low-redshift sources, simply the Hubble law.

An accurate distance measurement is also crucial for tests of modified theories of gravity aimed at constraining the speed of gravity~\cite{Belgacem:2017ihm,Iampieri:2024dul, Colangeli:2025bnb} and to distinguish between families of primordial BHs formed in the early universe and those formed from the first stars~\cite{Ng:2021sqn,Franciolini:2021xbq,Ng:2022agi,Ng:2022vbz,Franciolini:2023opt}.

All these examples highlight the utmost importance of measuring the luminosity distance $d_L$ and the inclination angle $\iota$ as accurately and precisely as possible with GWs.
However, a degeneracy between $d_L$ and $\iota$~\cite{Usman:2018imj} makes placing constraints over the two parameters difficult. The quasi indistinguishability of the polarization amplitudes in a broad region for $\iota\gtrsim0\degree$ is such that low-inclination systems are mistaken for closer, more inclined ones since the distance acts as a scale factor in the waveform. The issue is exacerbated by the selection effect for which face-on (face-off) binaries, namely ones with $\iota=0\degree \, (180\degree)$, are more likely to be detected~\cite{Gagnon-Hartman:2023soa}. Additionally, the inability of current instruments to detect the two polarizations with the same effectiveness~\cite{Usman:2018imj} might extend the problem to higher inclination angles~\cite{deSouza:2023gjv}. 

In principle, there are various ways to break this degeneracy: by using the measurements of $z$--$d_L$ and inclination from electromagnetic counterparts, as for GW170817~\cite{Tanvir_2017, Villar_2017}, by detecting higher-order modes~\cite{London_2018, Borhanian:2020vyr}, by measuring precession effects~\cite{Green:2020ptm, PhysRevLett.121.021303} or, for BNSs, through the binary Love relations between the progenitor tidal deformabilities~\cite{Xie:2022brn}.

In this work, we present a comprehensive exploration of the effects of the distance-inclination degeneracy on GW measurements using synthetic data in the zero-noise configuration, for a network of current and planned GW detectors, specifically for ET. 
We perform Bayesian inference using the public software \texttt{Bilby}~\cite{Ashton:2018jfp}.
We quantify the role of the network alignment factor, detector sensitivity, and waveform higher-order modes in mitigating or breaking the $d_L$--$\iota$ degeneracy, which is almost exact in the low-inclination limit.\footnote{
Reference~\cite{deSouza:2023gjv} performed a similar analysis for what concerns the alignment factor of the network. In addition to extending their analysis, here we also also investigate the role of the high-order modes and the inclination angles. 
}

The rest of this work is organized as follows. In Sec.\,\ref{phen:level1}, we introduce Bayesian inference techniques used to estimate CBC parameters. The degeneracy between $d_L$ and $\iota$ is discussed in Sec.\,\ref{degeneracy:level1}, while results are presented in Sec.\,\ref{sec:results} and divided into two sections corresponding to two distinct targets: the role of the network choice to detect the signal~\ref{subsec:align}, and the impact of high-order modes with ET~\ref{subsec:hm}. The paper ends with the conclusions of our study and with a discussion of its possible future extensions.

\section{\label{phen:level1}CBC Parameter Estimation}
Once a GW signal is detected, Bayesian inference~\cite{thrane2019introduction,Ashton:2018jfp} is carried out on the data to extract information about the physical parameters $\underline{\theta}$ of the source.  Namely, given information ``a priori'' that one assumes, and given a noise model of the GW detector(s), one may reconstruct the \textit{posterior distribution} $p(\underline\theta|s)$ of the source parameters by applying the Bayes theorem, conditioned by the detection of a total data stream
\begin{equation}
s(t)=h(t,\underline\theta)+n(t)\,,
\label{output_s}
\end{equation}
where $h(t,\underline\theta)$ is the GW signal, and $n(t)$ is the noise component due to the instrument(s). The Bayes theorem states that
\begin{equation}
    p(\underline\theta|s)=\frac{\mathcal{L}(s|\underline\theta)\pi(\underline\theta)}{\mathcal{Z}(s)}\,,
    \label{post}
\end{equation}
where:
\begin{itemize}

    \item $\mathcal{L}(s|\underline\theta)$ is the probability of measuring $s$ given the (source) parameters $\underline\theta$, and is known as the \textit{likelihood function}; the choice of the likelihood is linked to the noise model that is adopted;

    \item $\pi(\underline\theta)$ indicates the \textit{prior probability distribution} of having the set of parameters $\underline\theta$; it represents our knowledge about $\underline\theta$ before we make the measurement; and
    
    \item $\mathcal{Z}(s)$ is the \textit{evidence}, or marginal likelihood,
\begin{equation}
\mathcal{Z}(s)=\int\mathcal{L}(s|\underline\theta)\pi(\underline\theta)\underline{d\theta}\,,
\end{equation}
where the integral is intended over the full parameter space.
\end{itemize}

For CBC events, assuming quasi-circular orbits and working with post-Newtonian waveform models, the basic parameters that describe a binary system can be divided into two groups, for a total of 15.
\begin{itemize}
	\item The \textit{intrinsic parameters}, that depend only on the physical properties of the objects composing the source, and thus are independent of the observer: these are the masses $(m_1,m_2)$ and the spins of the compact objects $i=1,2$, written in their dimensionless form, $\bm{\chi}_i=c\mathbf{S}_i/Gm_i^2$, the effects of which come into play starting at 1.5-post-Newtonian order~\cite{Blanchet:2013haa}.
	\item The \textit{extrinsic parameters}, that describe the position of the source relative to the observer; they are the sky-location angles\footnote{The sky-location angles are often expressed in terms of the right ascension and declination ($\alpha,\delta$).} $(\theta, \phi)$ that localize the direction of the line of sight with respect to the detector reference frame, the polarization angle $\psi$ that defines the radiation frame, and the inclination angle and the luminosity distance, ($\iota, d_L$). To these one should add the coalescence time and phase ($t_c,\Phi_0$) as integration constants. 
\end{itemize}


	\section{ \label{degeneracy:level1}The distance--inclination degeneracy}
This section addresses the possibility of constraining the distance and inclination of observed GW signals emitted by compact binaries.  Starting from Ref.~\cite{Usman:2018imj}, we introduce the main reasons behind the distance--inclination correlation; in general, two parameters are said to be degenerate if a correlation exists, and this leads to an intrinsic difficulty in inferring them simultaneously. We then explain the scenarios to break the degeneracy that are used in this work; ultimately, the scope of our study is to carry out a survey on this degeneracy and explore specific configurations in the parameter space of the binary.	
	
\subsection{Origin of the degeneracy} \label{degeneracy:level2}
The measured GW signal expressed in terms of its two polarizations and of the detector responses is~\cite{Sathyaprakash:2009xs, Fairhurst:2017mvj},
	\begin{equation} \label{response_GW}
		h(t)=F_+(\theta,\phi,\psi)h_+(t)+F_{\times}(\theta, \phi, \psi)h_{\times}(t) \,,
	\end{equation}
	where $F_+$ and $F_{\times}$ are the antenna pattern functions
	\begin{equation}\label{complete_pattern_functions} \begin{split}
			F_+(\theta,\phi,\psi)= & \, \frac{1}{2}(1+\cos^2 \theta) \, \cos 2\phi \, \cos 2\psi \\
			& - \cos\theta \sin 2 \phi \, \sin 2\psi \\
			F_{\times}(\theta, \phi, \psi)= & \, \frac{1}{2}(1+\cos^2 \theta) \, \cos 2\phi \, \sin 2 \psi \\
			& + \cos\theta \sin 2 \phi \, \cos 2\psi \, .
	\end{split}\end{equation}
The two polarizations may be written, at Newtonian level (i.e. $0$-post-Newtonian) as:

\begin{equation}
\begin{cases}
h_{+}=\mathcal{A}_{+}\cos(\Phi(\tau))\\
h_{\times}=\mathcal{A}_{\times}\sin(\Phi(\tau))
\end{cases}\,,
\end{equation}
where $\tau=t-t_c$, $\Phi(\tau)$ is the phase to coalescence, and the Newtonian amplitudes written in a convenient form are
\begin{equation}
\begin{cases}
\mathcal{A}_+(d_L,\iota,\mathcal{M},\tau) = \frac{d_0}{d_L}\frac{1+\cos^2 \iota}{2} \\
\mathcal{A}_{\times}(d_L,\iota,\mathcal{M},\tau)= \frac{d_0}{d_L}\cos \iota \, 
\end{cases}\,,
\label{plus_cross_amplitudes}
\end{equation}
where $\mathcal{M}=(m_1m_2)^{3/5}/(m_1+m_2)^{1/5}$ is the chirp mass and $d_0=d_0(\mathcal{M},\tau)$ is a reference luminosity distance.

We refer to the wave as circularly and linearly polarized if $\iota=0\degree$ (or $\iota = 180\degree$) and $\iota=90\degree$, respectively. For a fixed distance $d_L$, identifying the inclination of the binary system using the polarizations of the GW requires being capable of distinguishing the contributions of the polarization amplitudes $\mathcal{A}_+$ and $\mathcal{A}_{\times}$.

\begin{figure} [t] 
\centering
\includegraphics[width=\columnwidth]{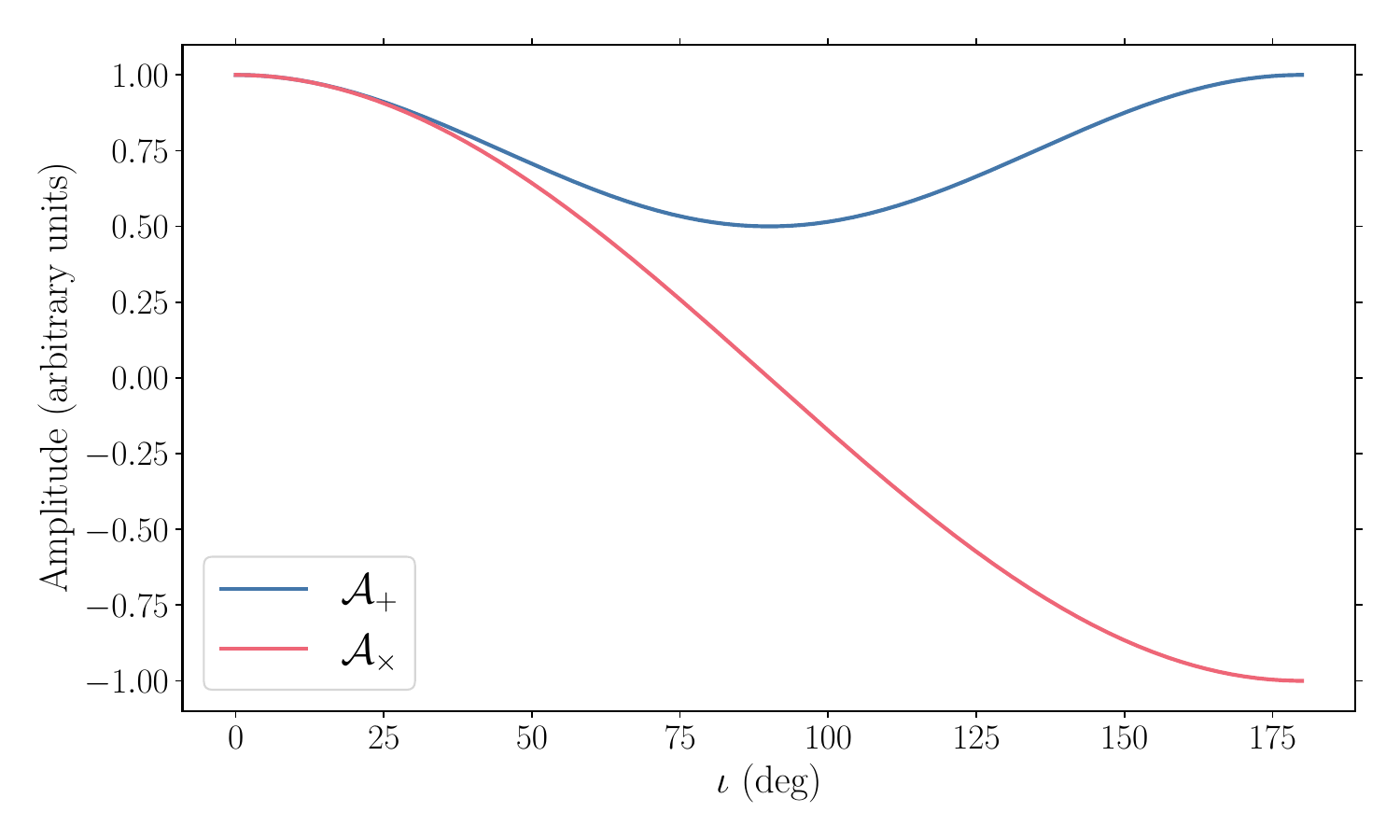}
\caption{Polarization amplitudes expressed in arbitrary units, referring to Eq.\,(\ref{plus_cross_amplitudes}), as a function of the inclination angle, for plus (blue) and cross (red) polarizations.}
\label{fig_amplitudeVSiota}  
\end{figure} 

Figure \ref{fig_amplitudeVSiota} shows clearly how for nearly face-on ($\iota\simeq 0 \degree$) and face-off ($\iota \simeq 180\degree$) binaries the polarization amplitudes are nearly equal in absolute value, and differ by only $\approx5\%$ at $\iota\approx45\degree$. 
This can be shown simply, by expanding $\cos\iota$ for $\iota\gtrsim 0\degree$ in the plus amplitude: 
	\begin{equation}
		\mathcal{A}_+ \propto \frac{1+\cos^2 \iota}{2}=1-\frac{\iota^2}{2}+\mathcal{O}(\iota^4) \simeq \cos \iota \propto \mathcal{A}_{\times}\,.
	\end{equation}
\emph{As a consequence, the two polarization amplitudes are hard to distinguish and the inclination angle cannot be measured.}
	
Further, Ref.~\cite{Usman:2018imj} reflects upon the impact of the strength of the signal relative to the noise, encoded in the optimal signal-to-noise ratio (SNR). The latter is given by
\begin{equation} \label{eq_opt_snr}
    \hat{\rho}=\sqrt{(h|h)}\,,
\end{equation}
where the scalar product between two signals $a(t)$ and $b(t)$ is defined as
\begin{equation} \label{inner_product}
    (a|b)=4\mathbf{Re}\int_{0}^{\infty}\frac{\tilde{a}(f)\tilde{b}^*(f)}{S(f)}\,,
\end{equation}
with a tilde ($\tilde{}$) denoting the Fourier transform, an asterisk (${}^*$) denoting the complex conjugate, and $S(f)$ being the power spectral density. We will refer to this quantity as either 
$\hat{\rho} $ or SNR interchangeably throughout this work. As noted in Ref.~\cite{Usman:2018imj}, the optimal SNR for small inclinations has the dependency
\begin{equation} 
\label{SNR_low_angles}
\hat{\rho}\left(\iota \sim 0\degree\right) \propto \frac{\cos \iota}{d_L}\,.
\end{equation}
\emph{This implies that detectors with limited reach in distance are biased to detecting signals from face-on/face-off systems, as these are the loudest ones at a fixed distance.}

\subsection{Scenarios to break the degeneracy relevant to this work}

\subsubsection{Alignment factor} \label{sec_alignment_factor}
A necessary condition for handling degeneracy is strongly linked to the ability of the network to get information from both polarization amplitudes. The ability of constraining both polarizations is quantified by the network \textit{alignment factor} [we refer the reader to Ref.~\cite{Klimenko:2006rh} for a complete derivation of this quantity]. Both the detector response --- Eq.\,\eqref{response_GW} --- and the log-likelihood ratio remain invariant under a rotation of the radiation frame by an angle $\psi$. If the network is composed of only one interferometer, the angle $\bar{\psi}$ that maximizes the plus-antenna pattern function of Eq.\,(\ref{complete_pattern_functions}) is given by
\begin{equation}
\frac{\partial}{\partial \psi}F_+(\theta, \phi, \psi)\Big{\vert}_{\psi=\bar{\psi}}=0\,,
\end{equation}
but with this choice, the cross-pattern function vanishes:
\begin{equation}
F_{\times}(\theta, \phi, \bar{\psi})=0\,.
\end{equation}
This implies that a single instrument cannot measure both polarizations.

To handle the case of a network composed of $N_{\text{det}}>1$ detectors, the starting point is the log-likelihood ratio \cite{Klimenko:2006rh}
\begin{equation}
\begin{split}
\lambda_{\text{LR}} & \equiv  \, \ln \left( \frac{\mathcal{L}(s|\underline\theta)}{\mathcal{L}(s|\text{noise})} \right)= \\ & = \sum_{k=1}^{N_{\text{det}}} \left( -\frac{1}{2} ( s_k - h_k, s_k - h_k )_k + \frac{1}{2} ( s_k, s_k )_k \right) ,
\end{split}
\end{equation}
where $(\cdot|\cdot)_k$ is the inner product defined in Eq.\,\eqref{inner_product} for the $k$-th instrument.  Expressing the total data stream, $s_k$, and response, $h_k$, of the $k$-detector as discrete frequency series, and introducing the \textit{network data vectors},
\begin{equation}
X_{+,\times}[f]=\sum_{k=1}^{N_{\text{det}}}\frac{F_{+,\times_k}s_k[f]}{S_k[f]} ,
\end{equation}
the log-likelihood ratio can be recast in a vectorial form,
\begin{align}
\lambda_{LR} &= \left( X_+^T \, \, X_\times^T \right) \begin{pmatrix} h_+^* \\ h_\times^* \end{pmatrix} -\\
&- \frac{1}{2} \sum_{f=0}^{f_{\text{max}}} \left( h_+[f] \, \, h_\times[f] \right) \mathbf{\mathcal{F}}[f] \begin{pmatrix} h_+^*[f] \\ h_\times^*[f] \end{pmatrix}\,,\nonumber
\end{align}
where $\mathcal{F}$ is the \textit{network response matrix}, which has general form
\begin{equation}
\mathcal{F}[f]= \left( \begin{array}{cc}
    \sum_{k=1}^{N_{\text{det}}} \frac{F_{+k}^2}{S_k[f]} & \sum_{k=1}^{N_{\text{det}}} \frac{F_{+k} F_{\times k}}{S_k[f]} \\[10pt]
    \sum_{k=1}^{N_{\text{det}}} \frac{F_{+k} F_{\times k}}{S_k[f]} & \sum_{k=1}^{N_{\text{det}}} \frac{F_{\times k}^2}{S_k[f]}
    \end{array} \right)\,.
\end{equation}
For each frequency bin, it is possible to perform a rotation of the radiation frame such that the network is maximally sensitive to the plus polarization. This corresponds to the transformation that diagonalizes the network response matrix, leading to what is referred to as the \textit{dominant polarization frame},
\begin{equation}
\begin{split}
    \mathcal{F}^{\,\text{dom}}[f] &= \mathcal{U}[f] \mathcal{F}[f] \mathcal{U}^T[f]  = \begin{pmatrix} g_{++}[f] & 0 \\ 0 & g_{\times \times}[f] \end{pmatrix} \\
    & = g_{++}[f] \begin{pmatrix} 1 & 0 \\ 0 & \alpha^2[f] \end{pmatrix},
    \label{dom_pol_frame}
\end{split}
\end{equation}
where $\mathcal{U}[f]$ is the rotation matrix for the frequency bin $f$, while the eigenvalues $g_{++}[f]$ and $g_{\times \times}[f]$ depend in a non-trivial way on the antenna patterns and the power spectral densities of the individual detectors. The \textit{alignment factor} $\alpha[f]\in [0,1]$, defined as
\begin{equation}
\alpha[f] = \sqrt{\frac{g_{\times \times}[f]}{g_{++}[f]}}\,,
\end{equation}
quantifies the access of the network to the two polarizations. This can be understood by noting that the log-likelihood ratio in the new frame can be split in two terms
\begin{equation}
    \lambda_{LR} \equiv \lambda_1 + \lambda_2,
\end{equation}
which have the form
\begin{equation}
    \lambda_1 = \sum_{f=1}^{f_{\text{max}}} \left( X_1[f] h_1^*[f] - \frac{g_{++}[f]}{2} \left|h_1[f]\right|^2 \right),
\end{equation}

\begin{equation}
    \lambda_2 = \sum_{f=1}^{f_{\text{max}}} \left( X_2[f] h_2^*[f] - \frac{g_{++}[f] \alpha^2[f]}{2} \left|h_2[f]\right|^2 \right),
\end{equation}
and where the subscripts $(1,2)$ represent the two polarizations $(+,\times)$ in the new frame,
\begin{equation}
    \begin{pmatrix}
        h_1[f] \\
        h_2[f]
    \end{pmatrix}
    = \mathcal{U}[f]
    \begin{pmatrix}
        h_+[f] \\
        h_\times[f]
    \end{pmatrix},
    \quad
    \begin{pmatrix}
        X_1[f] \\
        X_2[f]
    \end{pmatrix}
    = \mathcal{U}[f]
    \begin{pmatrix}
        X_+[f] \\
        X_\times[f] ,
    \end{pmatrix}
\end{equation}
again for each frequency bin. The two limiting cases are therefore $\alpha = 0$, where all the information comes from the plus polarization, as the term proportional to the square of the \textit{cross} polarization vanishes, and $\alpha = 1$, where the network is equally sensitive to both polarizations.

It is crucial to note that the alignment factor is determined by the source location and the relative orientation of the detectors within the network, as it depends on the individual antenna pattern functions. 
In the one-detector configuration, the alignment factor vanishes, and we have no information about the cross-polarization. This is also the case for a network composed of two L-shaped interferometers, where the arms are perfectly aligned and/or anti-aligned.
\begin{figure} [t] 
	\centering
	\includegraphics[width=\columnwidth]{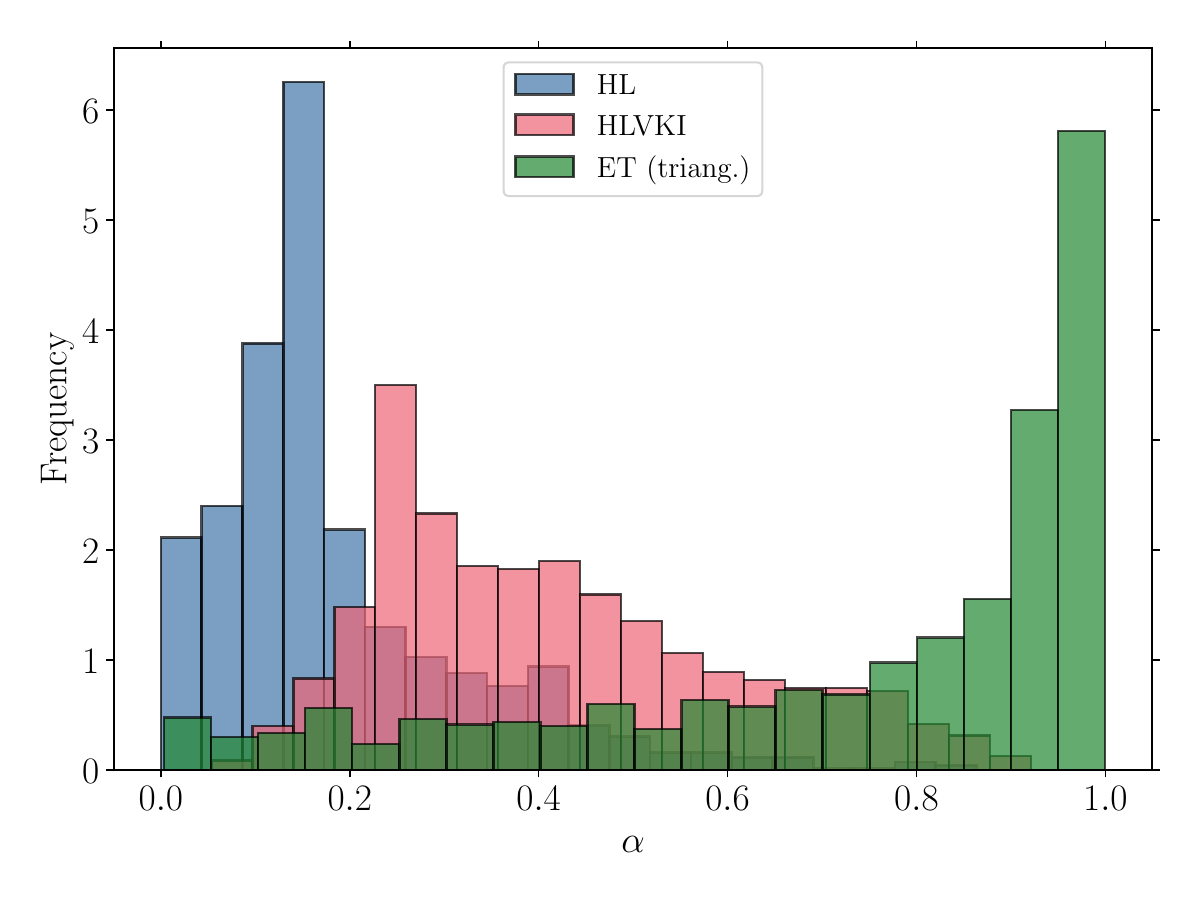} 
	\caption{Distribution of the alignment factor for three networks: Hanford–Livingston~(HL), Hanford-Livingston-Virgo-KAGRA-LIGO--India~(HLVKI), and Einstein Telescope~(ET) in the triangular design with 10~km long arms, obtained by varying the sky location. Each individual value of $\alpha$ corresponds to the average over frequency bins from 20~Hz to 1000~Hz.}
	\label{fig_alpha_hist}  
\end{figure} 
In Fig.\,\ref{fig_alpha_hist}, we show the distributions of this quantity for the three detector networks most frequently considered in this work: the first consisting of the two Advanced LIGO interferometers, located in Hanford and Livingston; the second including Advanced LIGO together with Virgo, KAGRA, and LIGO India; and the third composed of the three channels of the Einstein Telescope in its triangular design, with 10 km arms~\cite{Branchesi:2023mws}. These distributions are obtained by computing the alignment factor at different sky locations, varying the celestial coordinates accordingly. Each individual value of $\alpha$ is computed by averaging over frequency bins from 20~Hz to 1000~Hz. The mean values $\bar{\alpha}$ for the specific networks, are listed below for the reader's convenience.
\begin{itemize}
\item The Advanced LIGO network comprising Hanford and Livingston detectors, abbreviated as {\bf HL}:  $\Bar{\alpha}=0.23$;
\item The HL + Virgo + KAGRA + LIGO-India network, abbreviated as {\bf HLVKI}: $\Bar{\alpha}=0.46$;
\item Einstein Telescope\footnote{In this work, we consider only the triangular configuration, located in Sardinia.}, abbreviated as {\bf ET}: $\Bar{\alpha}=0.79$.
\end{itemize}
The low value of $\Bar{\alpha}$ for the two Advanced LIGO detectors is due to them having a nearly aligned arm and a nearly anti-aligned arm. On the other hand, ET is the most promising detector in terms of its ability to measure both waveform polarizations: its equilateral triangle-shaped design effectively corresponds to three misaligned detection channels~\cite{Punturo:2010zz, Branchesi:2023mws,Abac:2025saz}.

\subsubsection{High-order modes: the multipolar structure of gravitational radiation}

The observation of high-order modes in the waveform offers another possible way of breaking the degeneracy between $d_L$ and $\iota$. The general post-Newtonian expansion of the plus and cross polarizations is formally given by~\cite{Blanchet:2006zz, Blanchet:2013haa}:
\begin{align}
	h_{+,\times}(t)=&\frac{2G\mu x}{c^2 d_L} \, \underset{p=0}{\overset{+\infty}{\sum}}\left[x^{p/2}H_{+,\times}^{(p/2)}(m_1, m_2,\phi,\iota;t)\right]+ \nonumber\\
	&+\mathcal{O}\left(\frac{1}{d_L^2}\right) \,.
\end{align}
where $M=m_1+m_2$ is the total mass, $\mu = m_1m_2/M$ is the reduced mass, $x=\Bigl(GM \omega_s/c^3\Bigr)^{2/3}$, with $\omega_s$ the orbital angular velocity, and $\phi(t)$ is the phase of the waveform, expressed in turn as a post-Newtonian expansion.
$H_{+,\times}^{(p/2)}$ are instead the polarization amplitudes at the $p/2$-post-Newtonian level.\footnote{In general, the amplitudes are a function of other intrinsic parameters, such as spins and tidal deformabilities.} For instance, the 0.5-post-Newtonian amplitude corrections read~\cite{Maggiore:2007ulw}
\begin{equation}\begin{split}  
		h_+^{0.5}(t)=&\frac{G \nu}{4d_Lc^2} \, \delta m \, x^{3/2} \sin \iota \: \, \times \\
		& \times \, \left[(\cos^2 \iota+5)\cos \phi-9(1+\cos^2 \iota)\cos 3 \phi\right] \\ 
		h_{\times}^{0.5}(t)=&\frac{3G \nu}{4d_Lc^2} \, \delta m \, x^{3/2} \sin 2 \iota \, \left[\sin \phi-3\sin3\phi\right] \, ,
\end{split}\end{equation}
where $\delta m=|m_2-m_1|$ and $\nu=\mu/M$. 
The above corrections include a mass octupole and a current quadrupole which introduce terms depending on $\phi$ and $3\phi$, at variance with the mass quadrupole that depends on $2\phi$.
Although the amplitudes are equal and vanish for $\iota=0\degree$, for $\iota \gtrsim 0\degree$ the time evolutions are quite different. In particular, those amplitudes can be better distinguished as $\delta m$ increases, thus breaking the degeneracy that was evinced in Eq.\,(\ref{plus_cross_amplitudes}).

\begin{figure} [t] 
	\centering
	\includegraphics[width=\columnwidth]{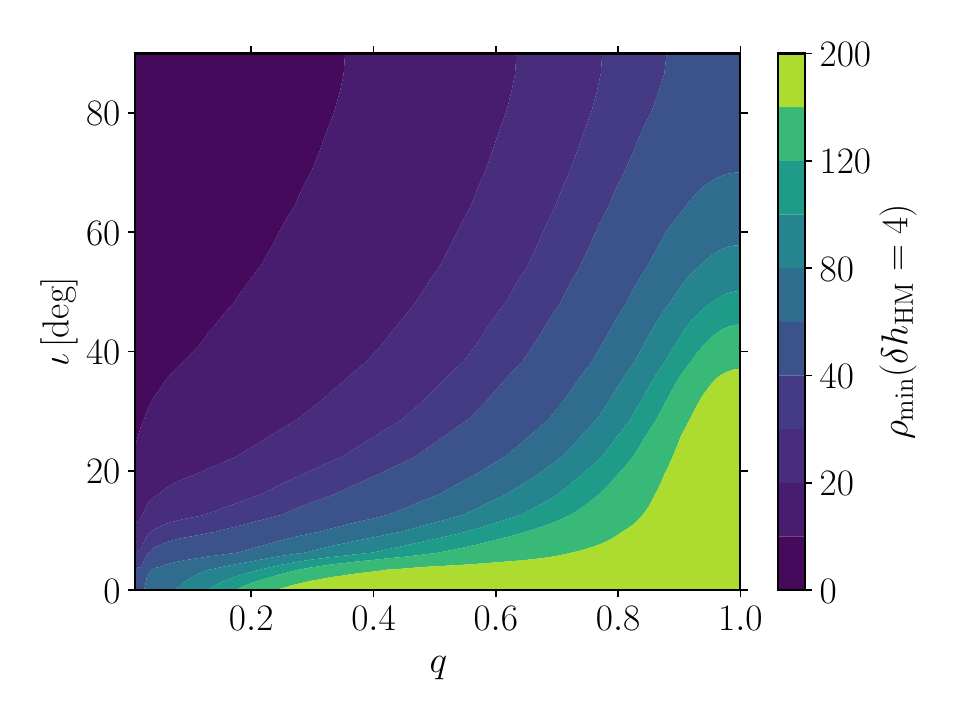} 
	\caption{The minimum optimal SNR required to obtain a \textit{distinguishability} between waveforms with and without subdominant modes included, according to Eq.\,(\ref{SNR_minimum}), as a function of mass ratio $q$ and inclination $\iota$. When the effect of higher-order modes is weaker, i.e., in the low-inclination and close to unity mass ratio region, a larger SNR is required to distinguish the waveforms.}
	\label{fig_distinguishability}  
\end{figure}

It is worth providing a qualitative behavior of the high-order terms over the mass ratio-inclination space ($q,\iota$), where $q=m_2/m_1$ with $m_2<m_1$. We first generated the dominant-multipole-only waveform $h_{22}$, selecting the $(2,\pm 2)$ modes, with \texttt{IMRPhenomXHM}~\cite{Pratten:2020ceb}, and then the complete version of the waveform, $h$, with all modes available in \texttt{IMRPhenomXHM} included. The \textit{overlap} between the two waveforms is defined as
\begin{equation} \label{match}
	O[h,h_{22}]= \, \underset{\{\Phi_0, t_c\}}{\max} \frac{\left( h|h_{22}\right)}{\sqrt{\left( h|h\right) \, \left( h_{22}|h_{22}\right)}} \, ,
\end{equation} 
and expresses their mutual compatibility as a normalized quantity. 
We define the \textit{distinguishability} of the high-order modes as~\cite{Lindblom:2008cm, Pannarale:2011pk, Baird:2012cu}
\begin{equation}
	\delta h_{\text{\rm HM}} \equiv \left( h-h_{22} | h-h_{22} \right) \simeq 2[1-O(h_{22},h)]\left( h|h \right) \, ,
\end{equation}
where the approximation is made in identifiying $\left( h_{22}|h_{22} \right) \simeq \left( h|h \right)$, which is the squared optimal SNR defined in Eq.\,(\ref{eq_opt_snr}) for the signal $h$. 
In Fig.\,\ref{fig_distinguishability} we report the minimum SNR that yields $\delta h_{\rm HM}=4$ (chosen as a reference value) as a function of $q$ and $\iota$, that is, we show
\begin{equation} \label{SNR_minimum}
	\hat{\rho}_{\text{min}}(\delta h_{\text{\rm HM}}=4) \simeq \sqrt{\frac{2}{1-O[h,h_{22}]}}
\end{equation}
as a function of mass ration and inclination.
This representative behaviour refers to a non-spinning binary with chirp mass $\mathcal{M}=14.8 \, M_{\odot}$ (for $q=0.1$, the component masses are $m_1=60 \, M_{\odot}$ and $m_2=6 \, M_{\odot}$) and the triangular ET sensitivity curve. The region that requires stronger signals to detect the subdominant modes is at low-inclinations and mass ratio values close to unity.

\section{\label{sec:results}Results for zero-noise injection simulations} 

\subsection{Setup}
In this section, we investigate strategies to break or mitigate the $d_L$--$\iota$ degeneracy by performing a zero-noise injection-recovery study using Bayesian inference simulations carried out with \texttt{Bilby}~\cite{Ashton:2018jfp}.
The baseline for our results is the set of simulations performed in Ref.~\cite{Usman:2018imj}, based on 4-parameter CBC Bayesian inference. The authors sampled over coalescence phase, polarization angle, luminosity distance and inclination, keeping fixed the coalescence time, the sky location, and the intrinsic parameters (with spins set to zero). The first generalization we bring is the inclusion of the mass intrinsic parameters among the sampled ones, albeit still maintaining the zero-spin configuration and the fixed sky location. 
In all the results that we present, we therefore neglect the effect of precession in the waveform, by imposing aligned spins injected to zero; we address the effect of precessing spins in the degeneracy in a future work~\cite{Crescimbeni_inprep}. 
Further, sampling over the sky-location parameters would indeed lead to larger uncertainties in distance and inclination measurements arising from varying detector sensitivities over the sky, while the focus of our investigating is the accuracy of measurements of distance and inclination in relation to the intrinsic degeneracy between the two quantities. Moreover, whenever a compact binary source with at least one neutron star is being considered, an electromagnetic counterpart observation could help in accurately localizing the source (possibly also constraining the inclination angle) thus reconducting the uncertainty problem back to the one of our interest~\cite{Maggiore:2024cwf}.

\begin{table}[t]
	\centering
		\begin{tabular}{@{\hspace{0.2cm}}c@{\hspace{0.6cm}}c@{\hspace{0.2cm}}}
		        \addlinespace[0.5em]
		        \toprule[1.1pt]
		        \toprule[1.1pt]
			Parameter &  Injected value \\
       			\midrule[0.8pt]
			$m_1$ & $1.40 \, M_{\odot}$ \\
			$m_2$ & $1.39 \, M_{\odot}$ \\
			$\mathbf{\chi}_1$ & $0$  \\ 
			$\mathbf{\chi}_2$ & $0$ \\
			$\alpha$ & $4^{\rm h} \, 11^{\rm m} \, 35^{\rm s}$  \\
			$\delta$ & $7.5\degree$ \\
			$\psi$ & $0\degree$ \\
			$\Phi_0$ & $74.5\degree$ \\
        			\bottomrule[1.1pt]
        			\bottomrule[1.1pt]
		\end{tabular}
	\caption{Injected parameters of the Bayesian inference simulations}
    \label{tab_parameters_1}
\end{table}

We focus on the marginalized joint  posterior distribution,
\begin{equation}
	p(d_L,\iota|s)=\int p(d_L, \iota, \bm{\xi}|s) \, d\bm{\xi} \, ,
\end{equation}
which is the probability distribution of Eq.\,\eqref{post} marginalized over all parameters except luminosity distance and inclination. This can be further marginalized to produce two single-parameter distributions.  The quality of recovering each parameter is encoded in the \emph{accuracy} that can be represented by a certain ``distance'' between the median of the 1-D marginalized distribution and the injected parameter, that is, the one characterizing the simulated source. Given the use of injections in zero-noise realizations, the accuracy of parameter recovery is influenced by the choice of prior distributions and by the marginalization over the remaining parameters\footnote{In a zero-noise injection-recovery simulation using the same waveform model for both injection and recovery --- as is always the case in our setup --- the injected parameters correspond to the maximum of the full $N$-dimensional likelihood. However, this is not necessarily true for the marginalized likelihoods. Moreover, when examining the marginalized posterior distributions, the influence of the prior can shift the location of the posterior peak away from the true injected values.}. These effects can be quantified as a ``distance'' between the median of the 1-D marginalized distribution and the injected parameter, that is, the one characterizing the simulated source. However, this distance should provide statistical information, since we want to know if the true parameter is contained within a certain credible interval. This also depends on the measurement \emph{precision}, which is reflected in the spread of the distribution. Given a parameter $\theta^i$, a good statistical distance can be defined as the absolute difference between the quantile order related to the injected value and $0.5$, that is, the quantile order of the median value. 
Given a simulation with injected parameter $\theta^i_{\text{inj}}$ and estimated median $\theta^i_{\text{med}}$, we thus define the following statistical distance:
\begin{equation} \label{statistical_distance}
	\delta_{\text{st}}(\theta^i)=| \, P(\theta^i_{\text{inj}}|s)-0.5 \, |=\left| \, \int_{\theta^i_{\text{med}}}^{\theta^i_{\text{inj}}} p(\theta^i|s) \, d\theta^i \,   \right| \,,
\end{equation}
where $p(\theta^i|s)$ is the posterior marginalized over all the parameters except $\theta_i$, and $P(\theta^i|s)$ is the cumulative of $p(\theta^i|s)$. This quantity approaches zero as the accuracy of the estimate increases. This quantity approaches zero as the estimated parameter gets closer to the injected one. Conversely, bad recoveries would push this quantity toward $0.5$. Whenever an injected parameter is at the boundary of the typical prior support for that variable, it does not  make sense to use this distance, since it will always return $0.5$. For instance, when injecting with an inclination angle of $0\degree$, a more indicative quantity might be the, e.g., the 90\% credible upper limit (or the lower limit, depending on the cases).

Another significant quantity, which however brings less statistical information, is the relative difference between injected and recovered parameter, identified by the median of the marginalized posterior distribution, namely $|\theta^i_{\text{inj}}-\theta^i_{\text{med}}|/\theta^i_{\text{inj}}$.  We will express this as a percentage.

\subsection{\label{subsec:align}Role of the alignment factor}

\begin{table}[t]
	\centering
		\begin{tabular}{@{\hspace{0.2cm}}c@{\hspace{0.6cm}}c@{\hspace{0.2cm}}}
		        \addlinespace[0.5em]
        			\toprule[1.1pt]
        			\toprule[1.1pt]
			Parameter &  Prior \\
			\midrule[0.8pt]
			$\mathcal{M}$ & Uniform in $[0.95\mathcal{M}_{\text{inj}},1.05\mathcal{M}_{\text{inj}}]$ \\ 
			$q$ & Uniform in $[0.5,\, 1]$ \\
			$\chi_{z, 1,2}$ & Aligned spin in [-1,1] \\
			$\cos \iota$ & Uniform in [$-1,1$] \\
			$d_L$ & Uniform-in-volume \\
			$\Phi_0$ & Uniform in [$0,2\pi$] rad \\
        			\bottomrule[1.1pt]
        			\bottomrule[1.1pt]
		\end{tabular}
    \caption{Priors in the most general configuration used. The ranges are specified when they remain unchanged in all the simulations.}
    \label{tab_priors_1}
\end{table}

For this first set of simulation, we make choices consistent with Ref.~\cite{Usman:2018imj}: we work with signals emitted by systems compatible with non-spinning BNSs, with (source-frame) masses $\sim 1.4 \, M_{\odot}$. The details of the properties of the injected signals and the priors used to recover them are reported in Tables \ref{tab_parameters_1} and \,\ref{tab_priors_1}, respectively.

We choose slightly different component masses for computational reasons, and we sample over (detector-frame) chirp mass and mass ratio, as these are estimated more directly from GW data. Chirp mass, in particular, is the best-estimated parameter and therefore, to make the simulations faster, we choose a uniform prior centered in its injected value and with a 10\% width around it. The injected mass ratio is close to unity, and the prior is taken to be uniform between $0.5$ and $1$. We provide priors for the (dimensionless) spin components aligned with the orbital angular momentum according to Ref.~\cite{Lange:2018pyp}. 
The extrinsic parameters listed in Table~\ref{tab_parameters_1} are unchanged throughout these simulations.

The injected inclination angle $\iota$ varies over the simulations: it is taken to give face-on, edge-on, or intermediate-inclination binaries. The prior associated to this quantity is uniform in the cosine of $\iota$. Finally, we use a uniform-in-volume prior for the luminosity distance, which is degenerate with the inclination angle; the support for this prior abundantly contains the injected distance, but it necessarily varies with the injected $d_L$ as this is modified to yield different values of optimal SNR.

\subsubsection{Two Advanced LIGO detectors} 

\begin{figure*} [t] 
	\centering
	\includegraphics[width=\columnwidth]{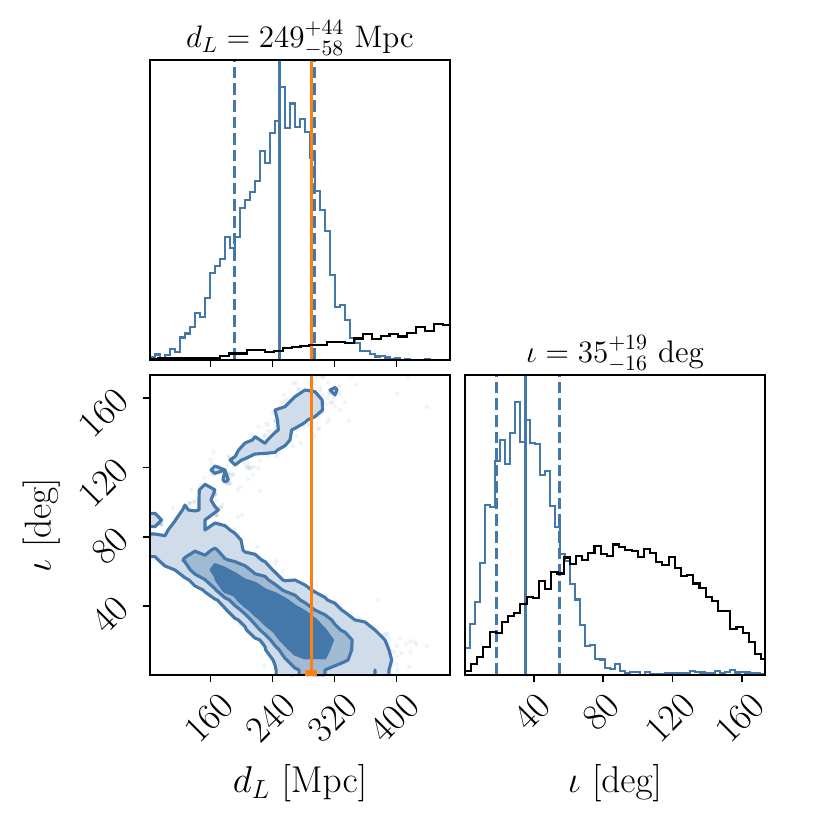} 
	\includegraphics[width=\columnwidth]{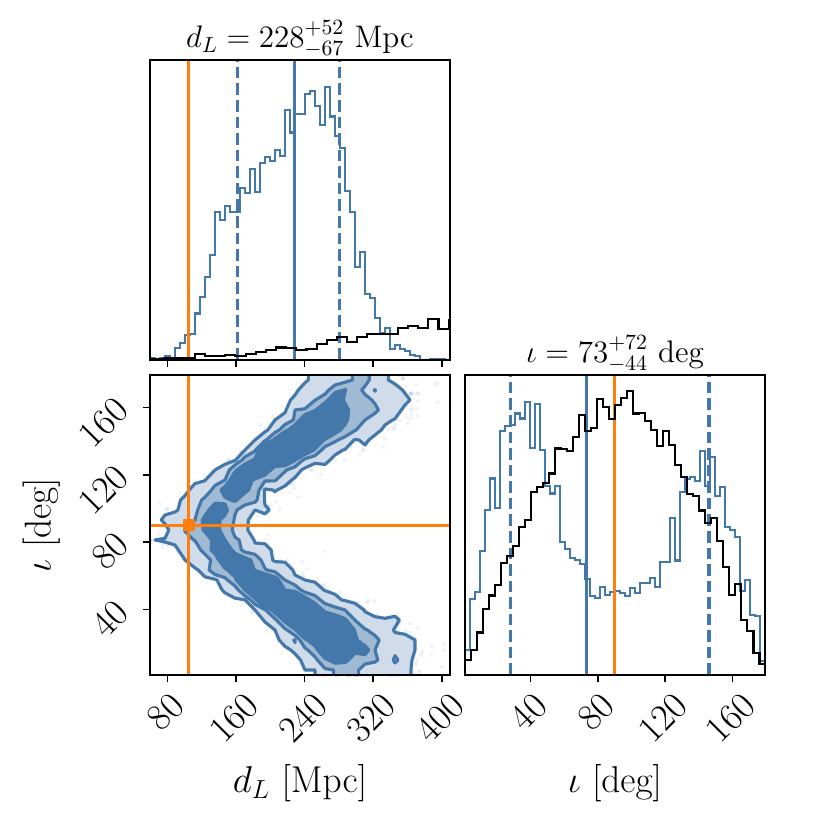}
 \includegraphics[width=\columnwidth]{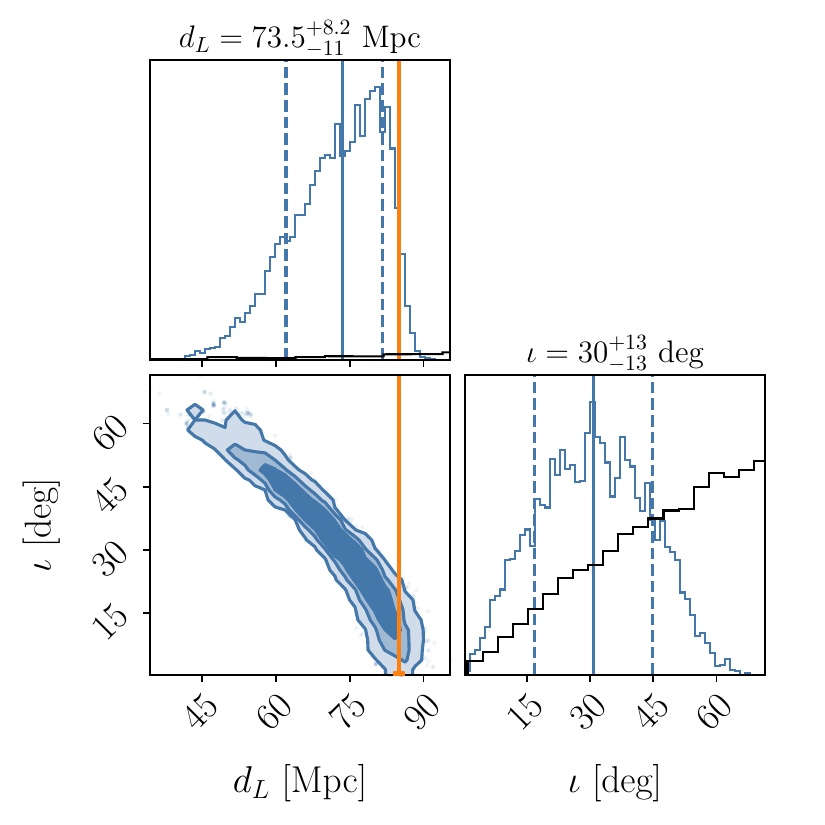} 
	\includegraphics[width=\columnwidth]{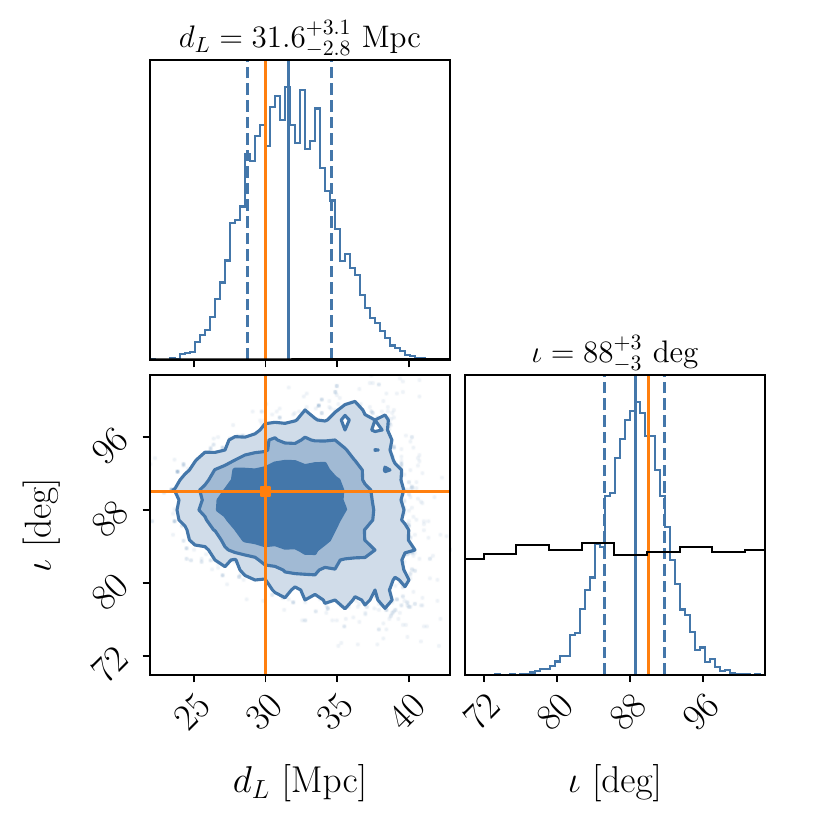} 
	\caption{Luminosity distance and inclination posterior distributions for face-on (left column) and edge-on (right column) simulations, with $\hat{\rho}=12$ (top row) and $\hat{\rho}=40$ (bottom row). The contours in the 2D joint posteriors mark the 68\%, 90\%, and 99\% credible regions. The vertical blue solid lines in the 1D posterior distributions indicate the median estimated values $d_L^{med}$ and $\iota^{med}$, while the dashed lines enclose the 68\% credible interval.  Finally, orange lines mark the injected values and in black we report the prior distribution.} 
	\label{fig_aLIGO_faceon_edgeon_12}  
\end{figure*} 

\begin{figure} [t] 
	\centering
	\includegraphics[width=\columnwidth]{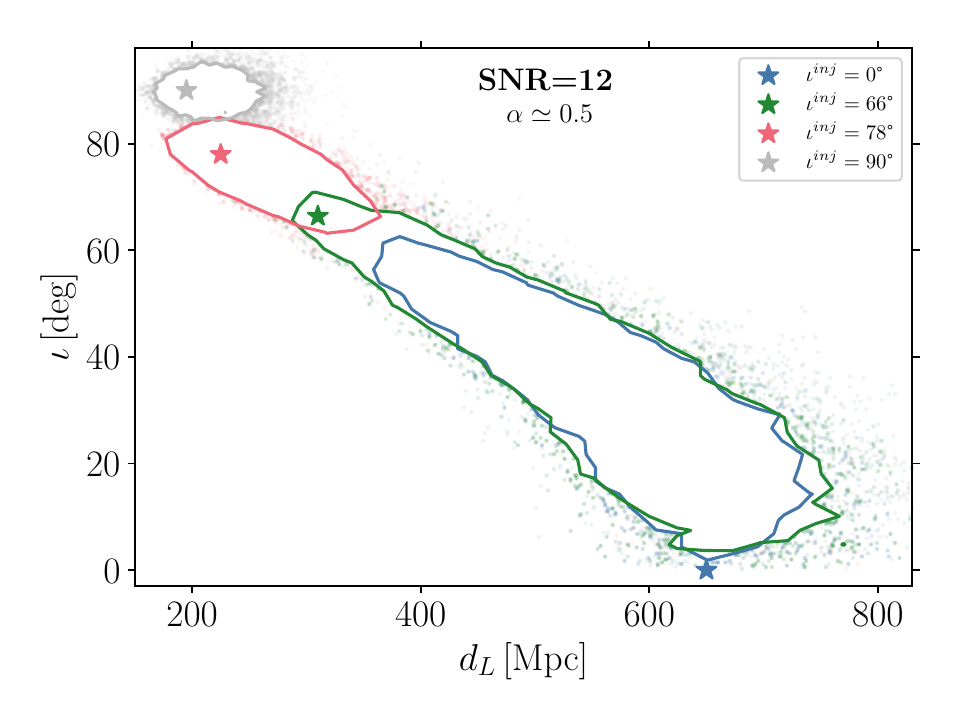} 
	\caption{90\% credible regions of the marginalized posteriors for different inclination angles; the stars mark the various injected values and each simulation is denoted by different colour.  The signals are assumed to be detected by the HLVKI, characterized at the chosen location by a frequency-averaged alignment factor $\alpha\simeq0.5$, with an optimal signal-to-noise ratio $\hat{\rho}=$12.}
	\label{fig_LVKI_contours_angles_12}  
\end{figure}

We first consider the HL network, composed by the two Advanced LIGO detectors and representative of low-$\alpha$ networks. In particular, at the location at which we performed this set of simulations, the value averaged over frequencies between 20 and 1000 Hz, is $\alpha\simeq 0.1$. We compare the face-on and edge-on configurations, and consider them at optimal SNR values of $12$ and $40$. The posterior distributions for $d_L$ and $\iota$ are represented in Fig.\,\ref{fig_aLIGO_faceon_edgeon_12}. 
We begin by considering the $\hat{\rho}=12$ cases (top row).  The stretched shape of the 2D distribution in the face-on configuration (left) is as expected.  Its $90\%$ containment region extends to $70\degree$, and it does not contain the injected value of $0\degree$. 
On the other hand, referring to Fig.\,\ref{fig_amplitudeVSiota}, the edge-on configuration should be the most favorable one in terms of distinguishing the two polarizations. 
However, in the $\hat{\rho}=12$ case the posterior tends to be more consistent with either a face-on or face-off system.  This is due to a combination of the low alignment factor and the moderately low SNR not allowing the cross amplitude to be observed. 
Because of this lack of information, the matched-filtering analysis tends to favour less inclined systems at larger distances. 
The origin of the insensitivity in this configuration can be explained as follows. A face-on binary emits circularly polarized radiation in the quadrupole approximation, and its waveform is described by a single overall amplitude and phase. Moreover, the related strain $h(t)$ depends on $\Phi_0$ and $\psi$ only in the combination $\Phi_0 \pm \psi$~\cite{Usman:2018imj}. 
Since the two amplitudes $\mathcal{A}_+$ and $\mathcal{A}_{\times}$ are (nearly) equal for (nearly) face-on binaries, terms such as $\cos(\Phi_0 \pm \psi)$ and $\sin(\Phi_0 \pm \psi)$ appear in the response and this additional degeneracy leads to a significantly larger parameter-space volume which is consistent with face-on/face-off, rather than edge-on systems.

We now compare these results to the ones for $\hat{\rho}=40$ (bottom row in Fig.\,\ref{fig_aLIGO_faceon_edgeon_12}). One can immediately check that the poor recovery for the two systems at $\hat{\rho}=12$ had different causes.
The face-on binary on the left maintains a degenerate profile which is smoother this time around, with a small enhancement in the precision\footnote{\label{note_precision_estimate}We can quantify the precision of the estimate as the ratio between $\Delta d_L$, that is the larger part of the 90\% credible interval, given the asymmetry of the distributions, and the median value $d_L^{\text{med}}$. We call it $\Delta d_L / d_L$ for simplicity.} $\Delta d_L/d_L$ from $36\%$ to $26\%$, and with the upper limit on the inclination angle dropping from $70\degree$ to $55\degree$. The right plot shows the edge-on case, and proves that $\hat{\rho}=40$ is enough to properly appreciate the difference between the two polarizations and to improve the recovery: the injected inclination is almost coincident with the median value, and is included in a 90\% credible interval $[84\degree,96\degree]$. As far as the distance is concerned, we find $\delta_{\text{st}}(d_L)=0.26$, so that the injection is inside the 68\% credible interval, with a moderate precision of $\Delta d_L/d_L=16\%$, which would further improve with a larger SNR.

\subsubsection{LIGO--Virgo--KAGRA--LIGO-India}

As we discussed in Sec.\,\ref{subsec:align}, having a network with more detectors is important to increase the network alignment factor. Indeed, the addition of Virgo~\cite{VIRGO:2014yos}, KAGRA~\cite{KAGRA:2018plz} and LIGO-India~\cite{LIGO-India-webpage} to the network (HLVKI) almost doubles its value (cfr. \ref{sec_alignment_factor}). At the chosen location, its value,averaged over frequencies between 20 and 1000 Hz, is approximately $\alpha\simeq0.5$.
We choose an optimal SNR $\hat{\rho}=12$ to make a direct comparison with the results of the previous section. Figure \ref{fig_LVKI_contours_angles_12} illustrates the joint posteriors as the injected inclination angle of the binary varies from $0\degree$ to $90\degree$.
The degeneracy remains present with an extended shape up to $\iota=66\degree$ while, for higher inclination angles, we witness an improvement. The statistical distance $\delta_{\text{st}}$ switches from 0.49 to 0.34, and, in contrast to the Advanced LIGO network, the edge-on system is fully recovered, with the median inclination that differs from the injected value by only 0.02\%, and $\Delta \iota / \iota=4\%$. The distance is estimated with a larger uncertainty of about 20\%. Hence, given the SNR and the alignment factor, there will be a threshold angle beyond which the estimates become acceptable within a chosen credible interval. On the other hand, if these two quantities are too low, as in the previous section with the Advanced LIGO detectors at $\hat{\rho}=$12, it may happen that even the most favorable configuration $\iota=90\degree$ remains unsolved, due to a lack in sensitivity that favours the more convenient low-inclination region.

\begin{figure} [t] 
	\centering
	\includegraphics[width=\columnwidth]{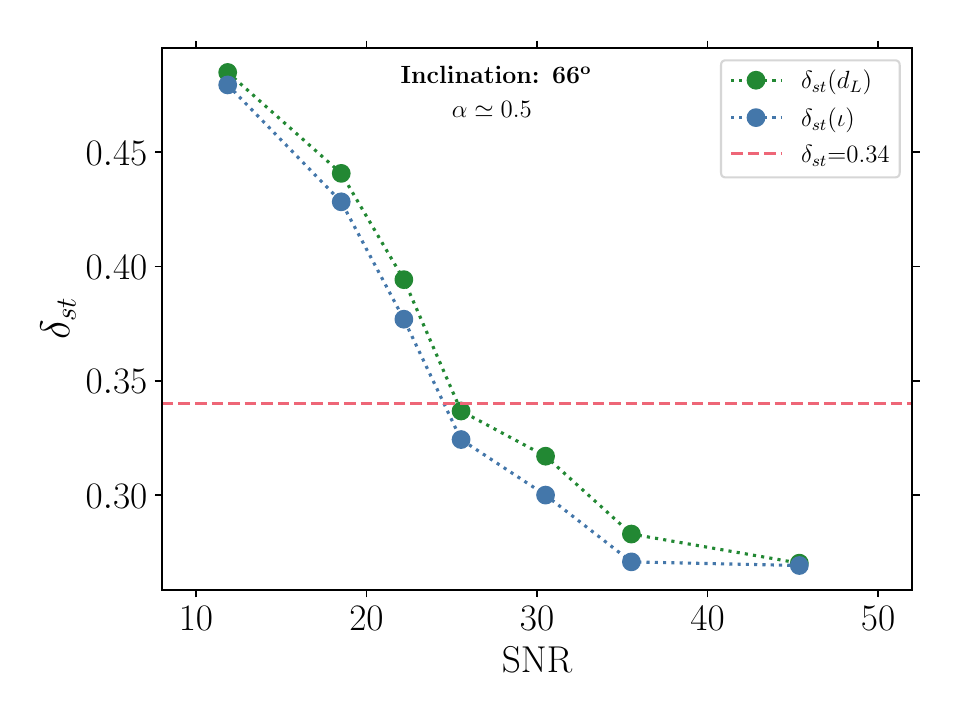}\\
	\includegraphics[width=\columnwidth]{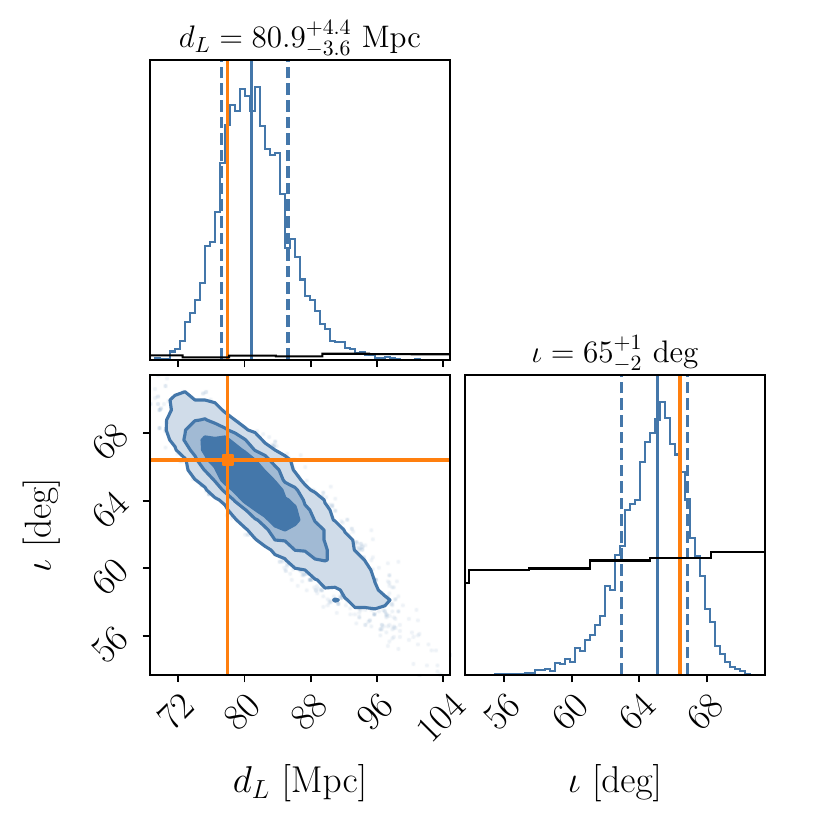}
	\caption{The top panel shows the statistical distance defined in Eq.\,(\ref{statistical_distance}) for both luminosity distance (green) and inclination (blue), as a function of SNR, for simulations carried out at a fixed inclination angle ($66\degree$). The dashed red line indicates the $\delta_{\text{st}}$ value that may be an acceptable threshold, since below this value the injection enters the 68\% credible interval. The lower panel is the joint $d_L$--$\iota$ posterior distribution for the rightmost point of the upper panel, characterized by $\hat{\rho}= 45$, with the same conventions as Fig.\,\ref{fig_aLIGO_faceon_edgeon_12}.}
	\label{fig_deltaVSsnr}  
\end{figure}

\subsubsection{The importance of signal-to-noise ratio and alignment factor}

\begin{figure} [t] 
	\centering
	\includegraphics[width=\columnwidth]{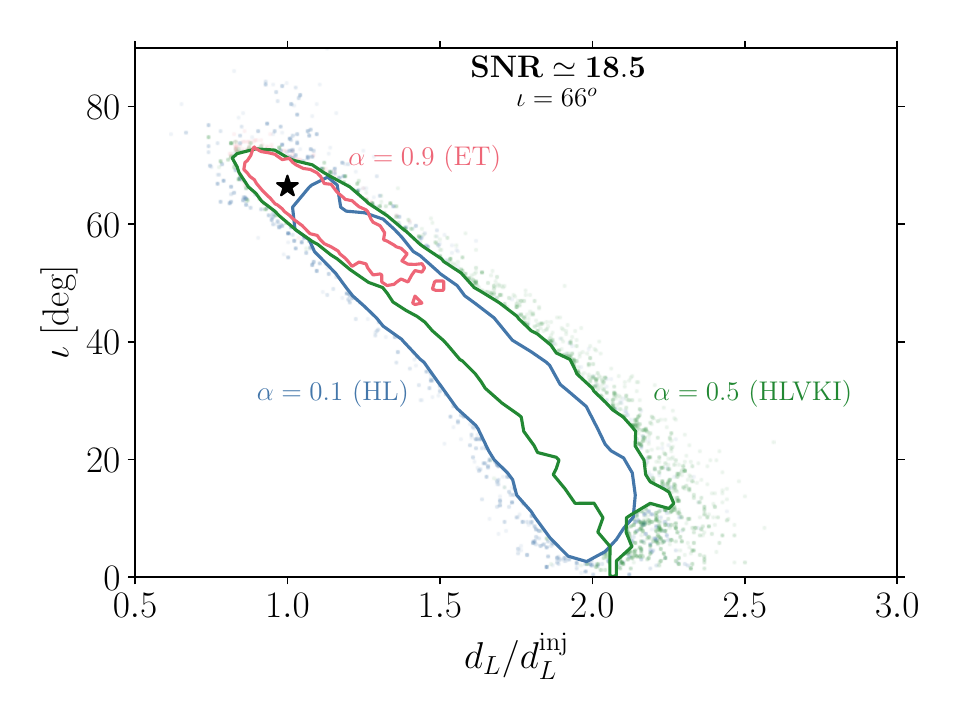} 
	\caption{90\% credible regions related to the same BNS-like system, with $\iota=66\degree$ and $\hat{\rho}=$18.5, detected by three different networks, each one associated to a different frequency-averaged alignment factor, reported in the plot. The distances are rescaled using the injected ones, adjusted for each network in such a way as to maintain the same SNR. The black star represents the injection.}
	\label{fig_contourVSalpha}  
\end{figure}  
The results presented so far suggest that the SNR is crucial for the recovery of distance and inclination for $\iota$ sufficiently different from zero, especially when the network is characterized by a low to moderate alignment factor. The Advanced LIGO detectors do not succeed in measuring the parameters even for an edge-on source if the SNR is too low. Still, the analysis improves drastically if the SNR is triplicated.  In Fig.\,\ref{fig_LVKI_contours_angles_12} we also saw that, at a given SNR, the degenerate behaviour persists up to angles $\sim 70\degree$ for a network with a larger alignment factor $\alpha \simeq 0.5$.  We now fix the angle to $\iota=66\degree$ and check that an increase in SNR leads to gradual improvement of the estimate. In the upper panel of Fig.\,\ref{fig_deltaVSsnr}, we plot the statistical distance $\delta_{\text{st}}$ defined in Eq.\,(\ref{statistical_distance}) as a function of the SNR, both for distance and inclination. 
Under these conditions, the threshold in SNR at which the injected value is within the 68\% credible interval is between 25 and 30. 
Moreover, the last point of the plot (at $\hat{\rho}\simeq45$) shows very similar accuracies for both parameters: referring to the joint marginalized posterior in the bottom panel of Fig.\,\ref{fig_deltaVSsnr}, we observe that the inclination is estimated with a precision $\Delta \iota/\iota \simeq 6\%$, while the precision in luminosity distance is 10\%. On the other hand, that is not the case for strongly degenerate configurations, when the angle is virtually not measured. 

\begin{table*}[t] 
	\centering
	\begin{tabular}{@{\hspace{0.2cm}}c@{\hspace{0.8cm}}c@{\hspace{0.8cm}}c@{\hspace{0.8cm}}c@{\hspace{0.8cm}}c@{\hspace{0.8cm}}c@{\hspace{0.8cm}}c@{\hspace{0.8cm}}c@{\hspace{0.2cm}}}
	        \addlinespace[0.5em]
		\toprule[1.1pt]
		\toprule[1.1pt]
		Network & $\alpha$ & $\delta_{\text{st}}(d_L)$ & $\Delta d_L/d_L$ & $|d_L^{\text{inj}}-d_L^{\text{med}}|/d_L^{\text{inj}}$ & $\delta_{\text{st}}(\iota)$ & $\Delta \iota/\iota$ & $|\iota^{\text{inj}}-\iota^{\text{med}}|/\iota^{\text{inj}}$   \\
		\midrule[0.8pt]
		HL & 0.1 & 0.48 & 33\% &65\% & 0.47 &  75\% & 48\%   \\
		HLVKI & 0.5 & 0.44 & 45\% & 45\% & 0.43 & 73\% & 25\% \\ 
		ET & 0.9 & 0.30 & 26\% & 10\% & 0.27 & 19\% & 5\% \\
		\bottomrule[1.1pt]
		\bottomrule[1.1pt]
	\end{tabular}
	\caption{A summary of the results obtained simulating the parameter estimation of a non-spinning, 66\degree-inclined BNS-like system detected at a fixed $\hat{\rho}=$18.5, with three different (mean) alignment factors, representative of the three different networks HL, HLVKI, and ET. 
    }	
    \label{tab_contourVSalpha}
\end{table*}

The other key factor in breaking the degeneracy is represented by the alignment factor $\alpha$, the impact of which has been widely discussed. We perform three simulations of the same system, with and inclination of 66\degree and detected at an optimal SNR $\hat{\rho}=18.5$ by three different networks: we add ET to the two networks already used so far, but at a different location with respect the one shown in Table \ref{tab_parameters_1}, chosen in such a way to obtain a frequency-averaged $\alpha \simeq 0.9$. This allows us to have a complete range from $0.1$ to $0.9$. Figure \ref{fig_contourVSalpha} reports the contours of the 68\% and 90\% credible regions.
The injection is inside the 90\% credible region in the low-$\bar\alpha$ case and enters the 68\% region when $\bar\alpha$ increases. The main differences between the three configurations can be observed in Table~\ref{tab_contourVSalpha}.
The precision of the estimates, shown in the third and sixth columns, improves for the inclination and seems to worsen for the distance, but this is not an issue, since we worked at fixed SNR.\footnote{This implies that the values of the distance were different such to have the same SNR.} On the other hand, the goodness of the recovery, described by the other columns, improves significantly, with the difference between the injected inclination and its median estimated value that drops from 48\% to 11\%. The values are slightly higher for the distance as discussed previously, but the enhancements are significant: from 65\% with Advanced LIGO HL to 26\% with ET.
We have thus proved that the configurations in which the inclination angle is quite far from zero are manageable with a combination of $\bar\alpha$ and $\hat{\rho}$ achievable by the upcoming observing runs. However, the recovery of less inclined systems remains a problem difficult to solve even for 3G detectors, unless the SNRs reach very high and rare values~\cite{deSouza:2023gjv}, or minor effects contribute to distinguishing the two polarization amplitudes and breaking the correlation. 

\subsection{\label{subsec:hm}Detecting high-order modes with ET}

The correlation between the luminosity distance and the inclination parameters mainly arises from the structure of the amplitudes [cfr., Eq.\,(\ref{plus_cross_amplitudes})], that are equal for $\iota=0\degree $ and differ by only $\sim1\%$ for $\iota=30\degree$. BNS and neutron star--black hole coalescing binaries could emit electromagnetic radiation, the detection of which could allow to properly measure the source redshift and luminosity distance~\cite{Hjorth:2017yza,Cantiello:2018ffy}. One may anyway wonder whether it is possible to significantly break such degeneracy with a GW-only detection. 

This section aims at investigating at what level the GW high-order modes break the degeneracy.  We will focus on face-on binaries, as we have observed that more inclined systems can be handled at relatively low SNRs with a moderate to high network alignment factor. 
Our Bayesian simulations in this context are carried out with ET in the triangular configuration~\cite{Branchesi:2023mws}. Later in this section we will discuss also different ET configurations.

\subsubsection{Injected parameters and priors} \label{sec:inj_parameters_2}
In the context of this section, we work with BBH systems as these can access low mass ratios which enhance the emission of GW high-order modes. Through this set of simulations, we fix the injected (detector-fame) total mass at $\sim 66 \, M_{\odot}$, in line with the total mass of GW150914~\cite{LIGOScientific:2016vlm}, and consider a relatively low mass ratio $q_{\rm inj}=0.1$, as well as a mass ratio closer to unity $q_{\rm inj}=0.8$. The prior on the (detector-frame) chirp mass is uniform, with a 10\%-width around the injected value; the prior on the mass ratio is also uniform: it is taken over the range $[0.3,1]$ for $q_{\text{inj}}=0.8$, and over the range $[0.02,0.8]$ for $q_{\text{inj}}=0.1$. 
The injected spins are still zero and are sampled under the assumption that they are aligned. The extrinsic parameters are roughly the same as for the first set of simulations, with two notable differences: the injected inclination angle is fixed to $0\degree$ and the prior on the luminosity distance is no longer uniform in volume. 
Since we work with binaries at cosmological distances, we need
to use astrophysical models. Thus, we implemented a prior uniform in comoving volume and in source-frame time, re-weighted with star-formation-rate information~\cite{Madau:2014bja, Fishbach:2018edt},
\begin{equation} \label{prior_SFR}
	p(z) \propto \frac{dV_c}{dz}\frac{1}{1+z}\psi(z) \, ,
\end{equation}
where the factor $(1+z)^{-1}$ converts from detector-frame to source-frame time, and $\psi(z)$ is the specific star-formation-rate,
\begin{equation}
	\psi(z)=0.015 \, \frac{(1+z)^{2.7}}{1+ \left[(1+z)/2.9\right]^{5.6}} \, M_{\odot} \, \text{yr}^{-1} \, \text{Mpc}^{-3} \, .
\end{equation}

\subsubsection{Low-mass-ratio binaries: an optimistic configuration}

%
\begin{figure} [!t] 
	\centering
	\includegraphics[width=\columnwidth]{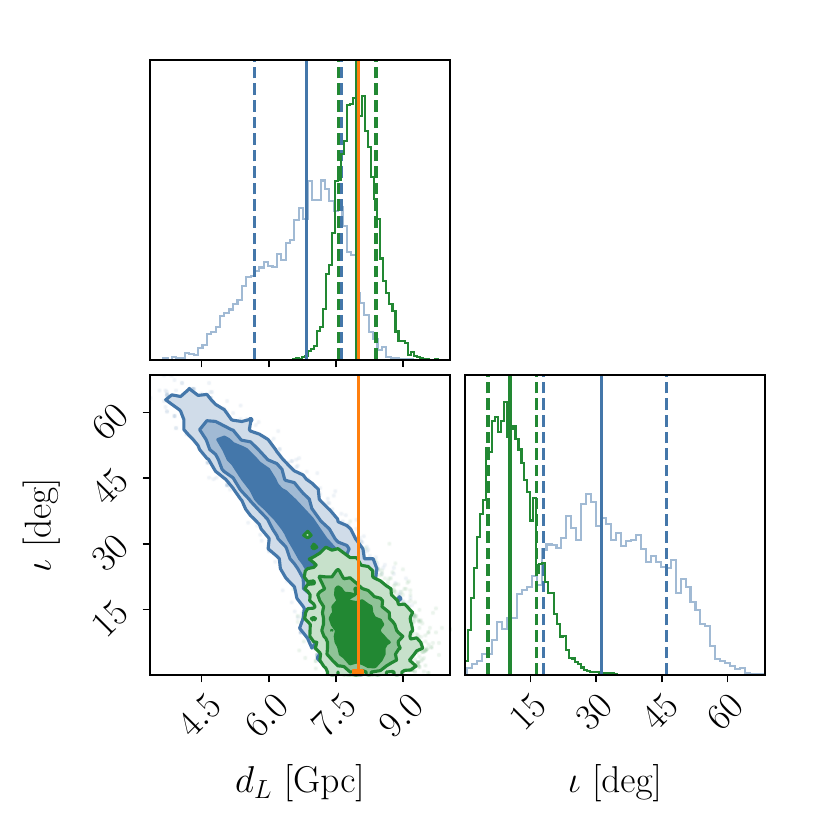}
	\caption{Joint $d_L$--$\iota$ posterior distributions related to the same face-on, $60 \, M_{\odot}$--$6 \, M_{\odot}$ binary detected at $\hat{\rho}=$20, optimally oriented toward the observer and detected with ET. The recoveries are performed with \texttt{IMRPhenomXP} (blue), and \texttt{IMRPhenomXHM} (green). In the 1D marginalized posterior plots, the vertical blue and green solid lines indicate the median estimated values, the dashed lines enclose the 68\% credible range and the orange lines mark the injected values, with a mark of the same color in the 2-D plots.}
	\label{ET_q01_20}  
\end{figure}

We start by considering a face-on system with $q_{\rm inj}=0.1$.  With a total mass of $66\,M_\odot$, the source for this configuration is a BBH with component masses $m_1=60 \, M_{\odot}$ and $m_2=6 \, M_{\odot}$ that are at the edge of the mass interval typical of stellar mass black holes. To witness the improvement that the inclusion of high-order modes brings to the posterior distribution, we inject signals with \texttt{IMRPhenomXHM}~\cite{Garcia-Quiros:2020qpx} at optimal SNR $\hat{\rho}=20$ and recover them either with \texttt{IMRPhenomXHM} itself, or with \texttt{IMRPhenomXP}~\cite{Pratten:2020ceb}, which does not include higher multipoles.  Despite not considering precession in this work, the choice of \texttt{IMRPhenomXP} is motivated by the need of a benchmark for future work on precessing effects~\cite{Crescimbeni_inprep}.

The resulting posterior distributions are shown in Fig.\,\ref{ET_q01_20}.
When high-order modes are not included in the approximant used for the recovery, despite ET's high alignment factor, the distribution maintains the stretched shape (the 90\% credible upper limit reaches 50\degree), due to the failure in discriminating the polarization amplitudes up to relatively large angles. When using high-order modes in the recovery stage, instead, the more inclined region is excluded by the inference.  This happens because the low mass ratio increases the amplitude of the high-order modes emitted by the source when this is viewed at high inclination angles: with this angle set to $0\degree$, an enhanced amplitude of these modes is incompatible with large values of $\iota$ when the injected signal is processed with an approximant that accounts for high-order modes.  In the face-on configuration, when the degeneracy tends to be broken with high-order modes, the inclination angle recovery is worse than the distance recovery, in contrast to what happened in the previous section: this is due to the fact that the injected value $\iota=0\degree$ is at the boundary of the support combined to the uniform-in-cosine prior that favors the edge-on region.  Yet, the 90\% upper limit of less than 20\degree is already quite constraining.

\begin{figure} [t] 
	\centering
	\includegraphics[width=\columnwidth]{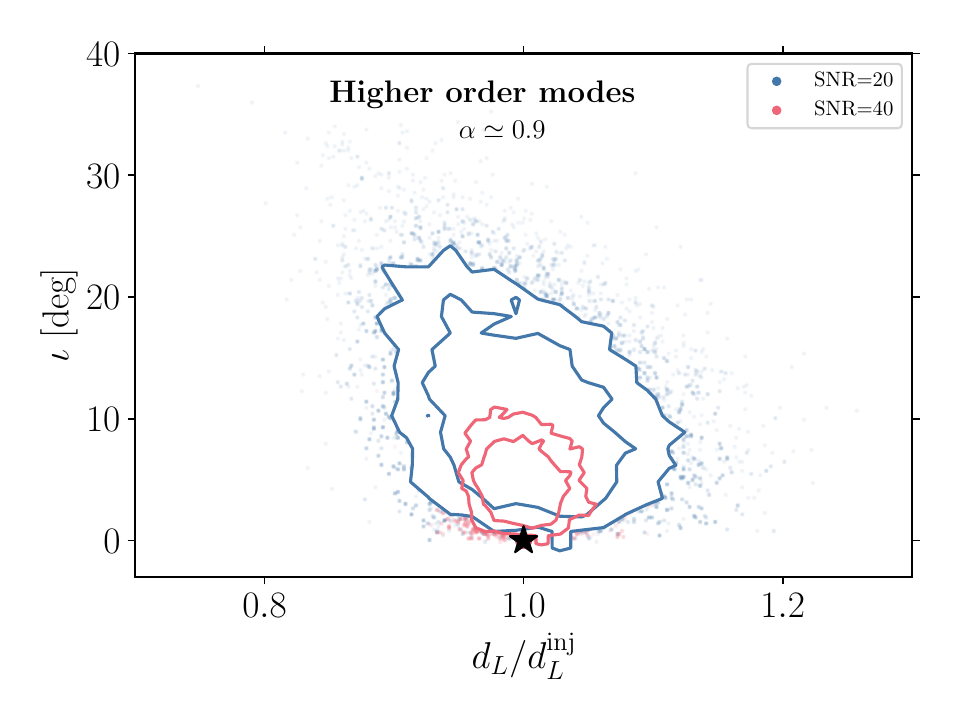}
	\caption{Joint $d_L$--$\iota$ posterior distributions related to a face-on,  $60\,M_{\odot}$--$6\,M_{\odot}$ binary detected with ET, including high-order modes. Contours indicate $68\%$, $90\%$, and $99\%$ credible regions. The results for $\hat{\rho}=20$ and $\hat{\rho}=40$ are shown in blue and green, respectively, while the black star represents the injection.}
	\label{fig_contours_ET_q01}  
\end{figure}

We now analyze the case of $\hat{\rho}=40$, when the SNR is twice the one seen so far, using \texttt{IMRPhenomXHM} for the recovery. The 2D marginalized posteriors for $d_L$ and $\iota$ are shown in Fig.\,\ref{fig_contours_ET_q01}.  As expected, the credible regions shrink by a factor $\sim (40/20)^2 = 4$.  The difference between injected and recovered distance is about $0.4\%$ for the higher SNR scenario, while the upper limit on the inclination angle drops to $\sim10\degree$ in this same case. The statistical distance $\delta_{\text{st}}(d_L)$ remains almost unchanged, while the precision is more than doubled. We summarize the results of these simulations in Table~\ref{tab_ET_q01}.

\subsubsection{\label{subsec:ET_design}The role of ET design for low-mass-ratio binaries}
We now focus on the impact of the ET design, the physical implications of which were carefully addressed in Ref.~\cite{Branchesi:2023mws}.
Specifically, we consider the same source system of $60M_{\odot}$ and $6M_{\odot}$, and simulate its recovery with three distinct ET designs:
\begin{enumerate}
    \item a triangular configuration with 10\,km arms (the current baseline ET geometry), located in Sardinia, Italy;
    \item a 2L configuration with 15\,km parallel arms, located in Sardinia and in the Netherlands;
    \item a 2L configuration with 15\,km arms, with a relative orientation of 45$\degree$, with the same locations of the previous configuration.
\end{enumerate}

\begin{table}[!t] 
\centering
\scalebox{0.8}{
\begin{tabular}{@{\hspace{0.2cm}}c@{\hspace{0.4cm}}c@{\hspace{0.4cm}}c@{\hspace{0.4cm}}c@{\hspace{0.4cm}}c@{\hspace{0.4cm}}c@{\hspace{0.2cm}}}
		        \addlinespace[0.5em]
		        \toprule[1.1pt]
		        \toprule[1.1pt]
Recovery model &   SNR & $\delta_{\text{st}}(d_L)$ & $\Delta d_L/d_L$ & $|d_L^{\text{inj}}-d_L^{\text{med}}|/d_L^{\text{inj}}$ &$\iota^{\text{upper}}$  \\
		        \midrule[0.8pt]
		\texttt{IMRPhenomXP} & 20 & 0.45 &25\% & 17\% &53\degree \\
	        \addlinespace[0.5em]
		\multirow{2}{*}{\texttt{IMRPhenomXHM}} & 20 & 0.05 & 9.5\%& 0.6\% &21\degree \\ 
		&  40 & 0.07 & 4\% & 0.4\% &10\degree \\
		        \bottomrule[1.1pt]
		        \bottomrule[1.1pt]
\end{tabular}
}
\caption{A summary of the results obtained simulating the parameter estimation of a non-spinning face-on binary, detected by ET. The injections are made with \texttt{IMRPhenomXHM}, while recoveries are performed with both \texttt{IMRPhenomXHM} and \texttt{IMRPhenomXP}. The 95\% upper limit on the inclination angle is taken.} 
\label{tab_ET_q01}
\end{table}

\begin{figure*}[!t]
    \centering
    \includegraphics[width=\columnwidth]{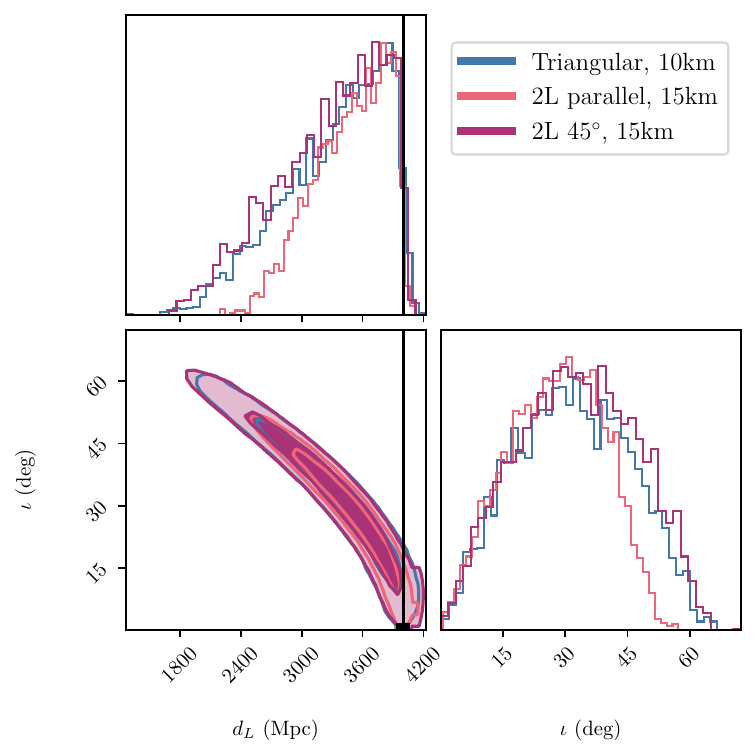}
    \includegraphics[width=\columnwidth]{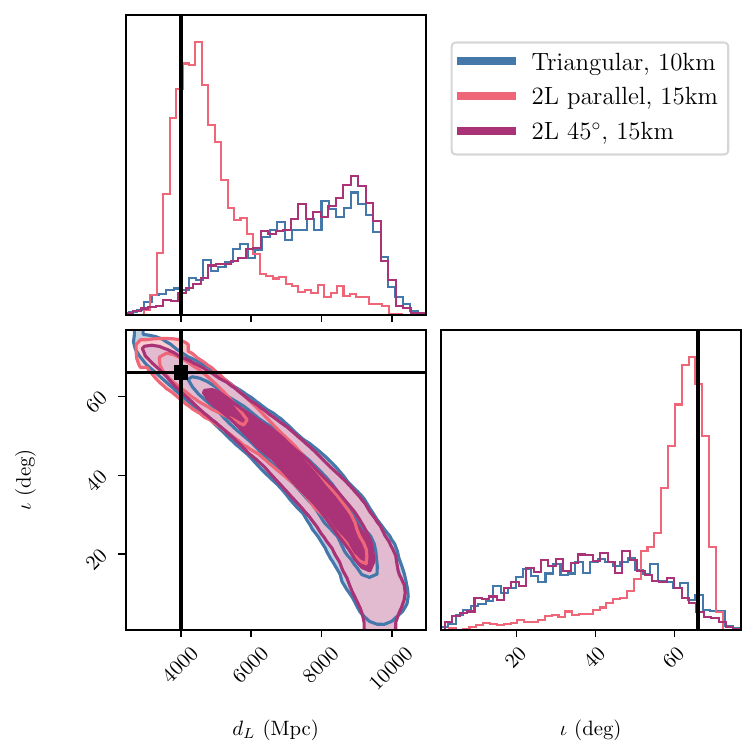}
    \includegraphics[width=\columnwidth]{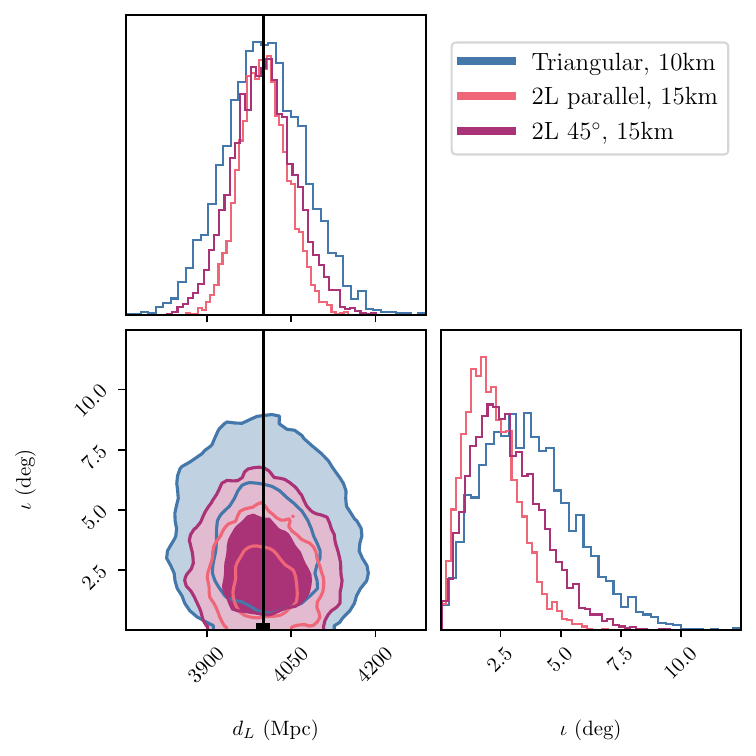}
    \includegraphics[width=\columnwidth]{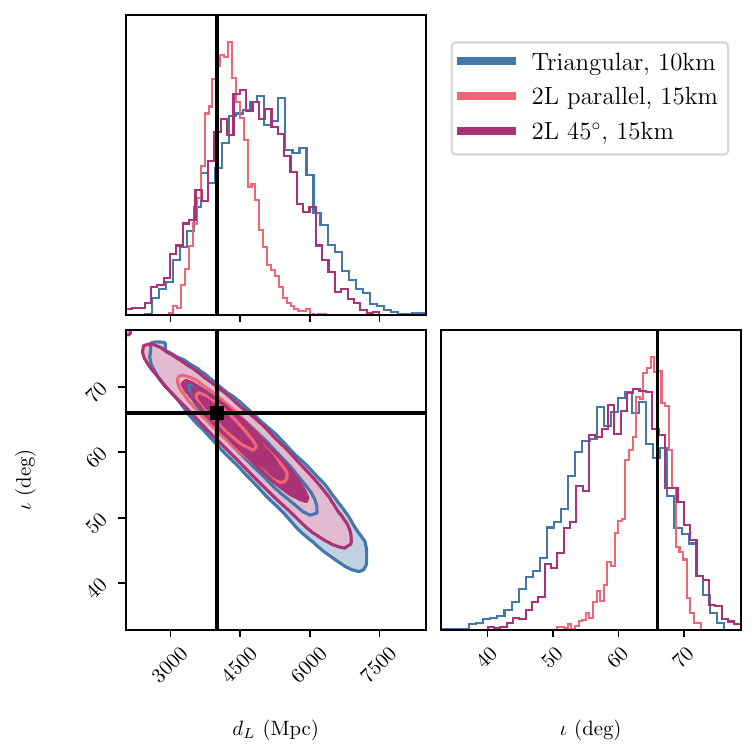}
    \caption{Joint $d_L$--$\iota$ posterior distributions for different ET designs, with injected $\iota=0\degree$ (left panels) and $\iota=66\degree$ (right panels); the recovery in the top and bottom panels is performed without and with higher-order modes, respectively. The black lines mark the injected values. The injected luminosity distance is fixed to 4\,Gpc, while the SNR values are reported in Table~\ref{tab:snr_ET_iota_design}.}
    \label{fig:corner_designs}
\end{figure*}

For each design, we set the inclination angle of the simulated source to two different values: $\iota\in\{0\degree, 66\degree\}$.  In both cases, we perform a recovery with and without high-order modes and the distance and sky location are kept unchanged. The corresponding optimal SNR values are reported in Table~\ref{tab:snr_ET_iota_design}.

The $d_L$--$\iota$ corner plots of the inference are shown in Fig.\,\ref{fig:corner_designs}.  Unsurprisingly, given its higher optimal SNR values, the 2L configuration with parallel arms is the one that yields the results with the lowest uncertainties. However, the different designs give comparable posteriors and thus for the analyses that follow we will assume ET within its official triangular design. We leave a more detailed analysis of the $d_L$--$\iota$ degeneracy with different ET designs for future work.

\subsubsection{Next-to-unity mass ratio binaries I: a more challenging configuration}
%
\begin{table}[t]
\begin{tabular}{@{\hspace{0.2cm}}c@{\hspace{0.6cm}}c@{\hspace{0.6cm}}c@{\hspace{0.6cm}}c@{\hspace{0.2cm}}}
		\addlinespace[0.5em]
        		\toprule[1.1pt]
        		\toprule[1.1pt]
		                   & Triangular & 2L parallel & 2L 45$\degree$ \\
		                	  & 10\,km & 15\,km & 15\,km \\
                 \midrule[0.8pt]
		$\iota=0\degree$   & 55               & 95                                     & 72                    \\
		$\iota=66\degree$ & 24               & 42                                     & 31                    \\
		\bottomrule[1.1pt]
		\bottomrule[1.1pt]
\end{tabular}
\caption{Optimal SNRs for a BBH source with (detector-fame) total mass $66M_{\odot}$ and mass ratio $0.1$, detected with different ET designs. The inclination angle is $0\degree$ and $66\degree$ in the first and second row, respectively. In both the cases, the injected distance is 4\,Gpc and the sky location is unchanged.}
\label{tab:snr_ET_iota_design}
\end{table}
We now switch to injected signals with mass ratio to $q_{\rm inj}=0.8$. The main reasons of interest in this configuration are two.
\begin{enumerate}
\item The majority of the events of the first three LIGO-Virgo-KAGRA observing runs are characterized by mostly quasi-symmetric binaries~\cite{LIGOScientific:2018mvr, LIGOScientific:2020ibl, LIGOScientific:2021djp}.
\item This is a value compatible with both BBH and BNS systems, and the results of the two cases can be compared.
\end{enumerate}

\begin{figure} [!t] 
	\centering
	\includegraphics[width=\columnwidth]{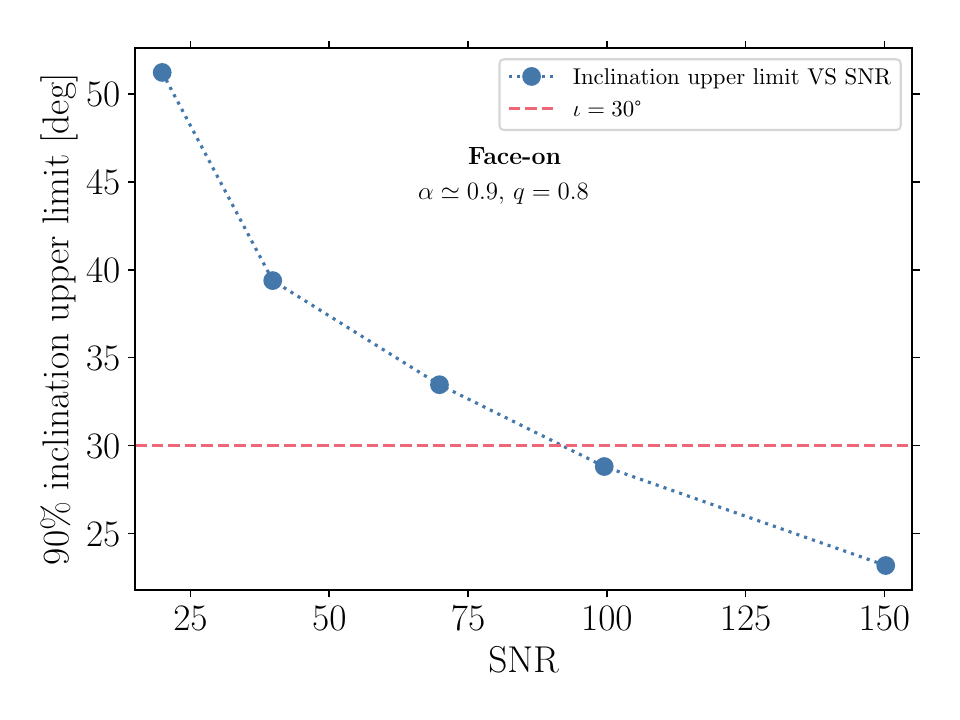}
	\includegraphics[width=\columnwidth]{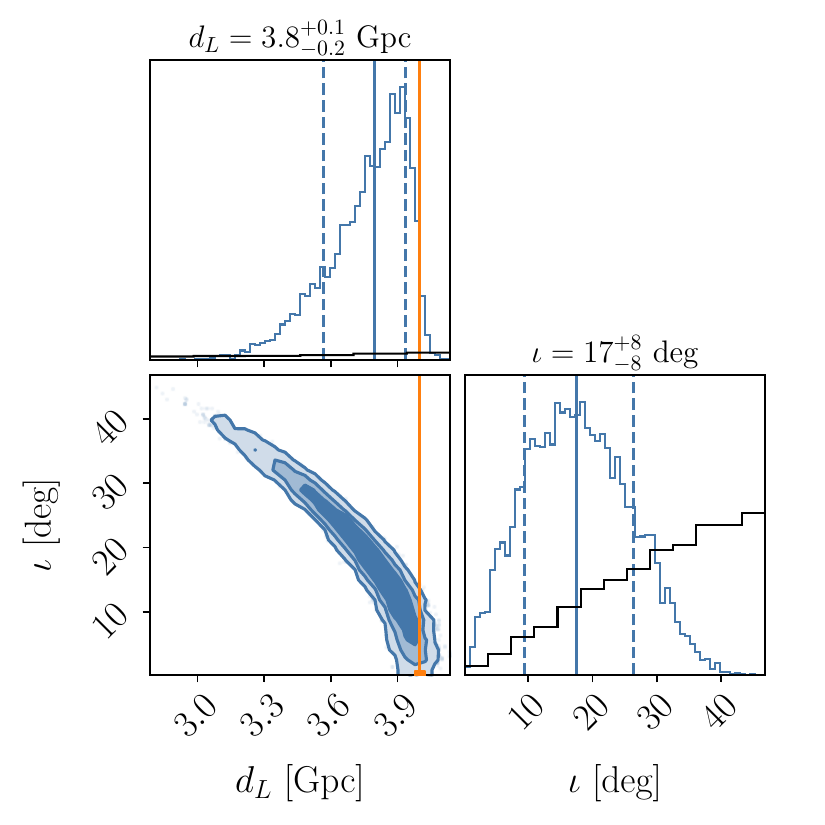}
	\caption{Upper panel: behavior of the 90\% upper limit on $\iota$ as a function of optimal SNR with ET. The parameter estimations are related to a face-on BBH system with component masses of $29\, M_{\odot}$ and $36 \, M_{\odot}$, with injections and recoveries performed with \texttt{IMRPhenomXHM}. The red-dashed line marks the $30\degree$ inclination, which qualitatively separates low-inclined from moderately-inclined binaries. Lower panel: joint $d_L$--$\iota$ posterior distributions related to the representative case with $\hat{\rho}=$100, with contours indicating $68\%$, $90\%$, and $95\%$ credible regions. The remaining conventions are the same as in Fig.\ref{fig_aLIGO_faceon_edgeon_12}.}
	\label{fig_ET_upperVSsnr_XHM_XHM_q08}  
\end{figure}

We begin by taking face-on BBHs with injected (detector-frame) masses of $36\, M_{\odot}$ and $29 \, M_{\odot}$, close to the GW150914 values; we also perform a simulation at the same mass ratio but with neutron-star-compatible masses in order to confirm the similarity of the outcomes.
For the BBHs, we varied the SNR from 20 to 150. As expected, the impact of high-order modes on the recovery is reduced, given that they are suppressed in the injected signal by the quasi-symmetry of the masses and therefore higher SNRs are appreciate them. The trend of the 90\% credible upper limit on the inclination angle as a function of the optimal SNR is reported in the upper panel of Fig.\,\ref{fig_ET_upperVSsnr_XHM_XHM_q08}, while the lower panel is the posterior distribution for the specific simulation with an optimal SNR of 100. The upper limit value on the inclination angle for $\hat{\rho}=150$ is almost the same as the one estimated at an SNR of 20 but with $q_{\rm inj}=0.1$ (see Table~\ref{tab_ET_q01}). The percentage difference between injected distance and recovered median value passes from 17\% to 3\%. However, the posterior distribution over the distance, with increasing SNR, does not tend to a normal distribution, but it shrinks asymmetrically toward the injected value that remains out of the 68\% credible interval, as shown in the lower panel of Fig.\,\ref{fig_ET_upperVSsnr_XHM_XHM_q08}. For this reason, the statistical distance $\delta_{\text{st}}(d_L)$ does not improve.

\begin{figure} [t] 
	\centering
	\includegraphics[width=\columnwidth]{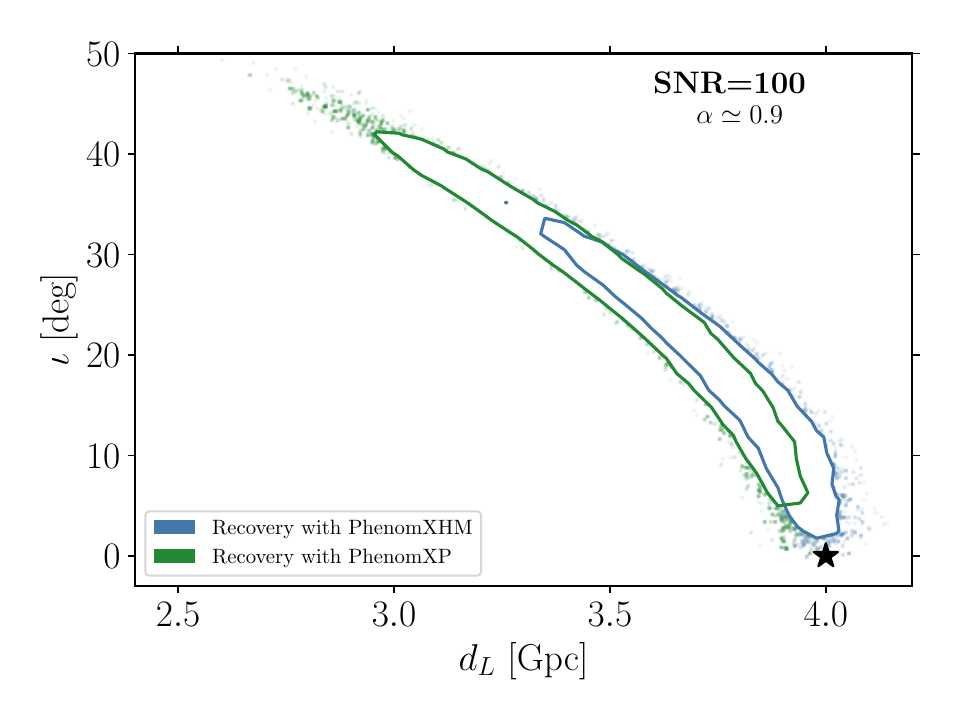}
	\caption{90\% luminosity distance--inclination angle credible regions for a face-on binary with masses $3\, M_{\odot}$ and $29 \, M_{\odot}$, detected by ET at an SNR of 100. The signal was injected with \texttt{IMRPhenomXHM} (indicated with the black star), and the recovery with both \texttt{IMRPhenomXP} and \texttt{IMRPhenomXHM}, indicated respectively in green and in blue.}
	\label{fig_ET_comparison_HM_XP_q08}  
\end{figure}

Figure \ref{fig_ET_comparison_HM_XP_q08} reports the difference induced by the usage of \texttt{IMRPhenomXHM} or \texttt{IMRPhenomXP} for the recovery, to emphasize that higher-order modes remain important, but their impact is limited by the high mass ratio. The 90\% contours are related to an SNR $\hat{\rho}=100$, and the key differences are summarized in Table~\ref{tab_ET_q08}.
This table shows that considering high-order modes remains beneficial for both the goodness of the recovery and the precision, but the comparison with Table~\ref{tab_ET_q01} highlights that a lower mass ratio is crucial.

Moving to BNS systems, we considered masses of $1.4\, M_{\odot}$ and $1.1 \, M_{\odot}$, maintaining the same optimal SNR $\hat{\rho}=100$, that for ET corresponds to an injected luminosity distance of $\sim 200$ Mpc. We again used the two different approximants for the recoveries in two separate runs to ensure consistency with the previous test. We do not report the marginalized posteriors, which are analogous to the ones in Fig.\,\ref{fig_ET_upperVSsnr_XHM_XHM_q08}, but we rather present Table~\ref{tab_ET_q08} to be compared to Table~\ref{tab_ET_BNS_q08}.

\begin{figure}[t] 
	\centering
	\includegraphics[width=\columnwidth]{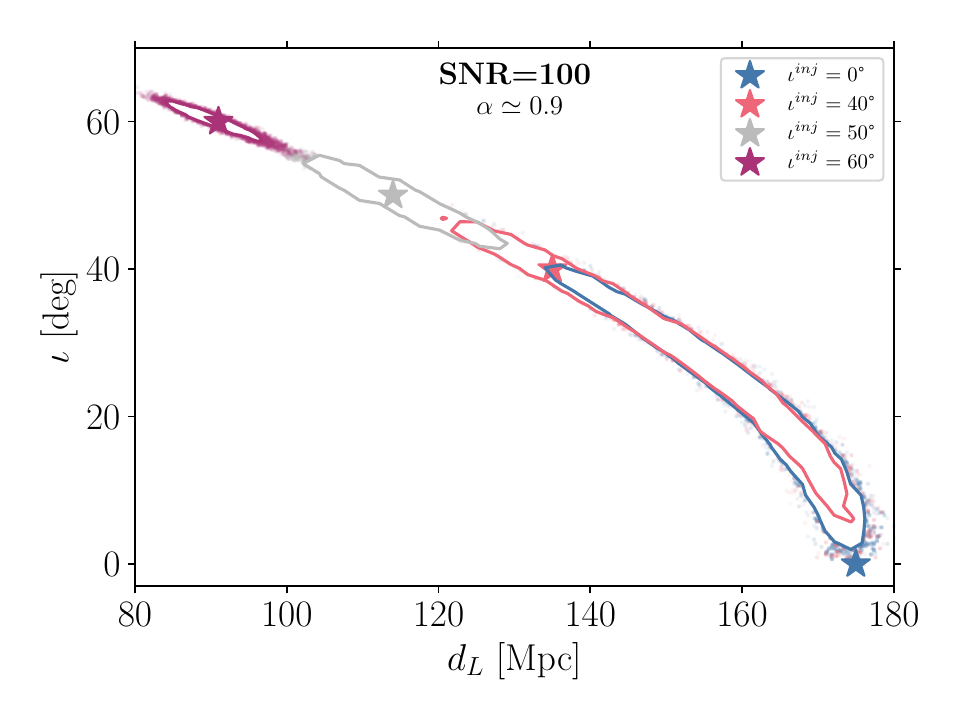}
	\caption{90\% luminosity distance--inclination angle credible regions for a face-on, BNS system with masses $1.4\, M_{\odot}$ and $1.1 \, M_{\odot}$, detected by ET at an SNR of 100, for different injected inclination angles. The high-order modes are included since \texttt{IMRPhenomXHM} is employed (a similar result holds for \texttt{IMRPhenomXP}).}
	\label{fig_contours_ET_BNS_100}  
\end{figure}

We conclude this section by presenting a result similar to the one of Fig.\,\ref{fig_LVKI_contours_angles_12}: we simulate the same BNS system but varying the inclination angle, while keeping the optimal SNR fix at 100. 
The joint posterior distributions are reported in Fig.\,\ref{fig_contours_ET_BNS_100}: it is clear that, within this sensitivity, we will be able to constrain the two parameters of interest for sources of this kind with inclination angles greater than or equal to $\sim 50\degree$.

\begin{table}[t] 
	\centering
	\scalebox{0.9}{
	\begin{tabular}{@{\hspace{0.2cm}}c@{\hspace{0.4cm}}c@{\hspace{0.4cm}}c@{\hspace{0.4cm}}c@{\hspace{0.4cm}}c@{\hspace{0.2cm}}}
		\addlinespace[0.5em]
		\toprule[1.1pt]
		\toprule[1.1pt]
		Recovery model &  $\delta_{\text{st}}(d_L)$ & $\Delta d_L/d_L$ & $|d_L^{\text{inj}}-d_L^{\text{med}}|/d_L^{\text{inj}}$ &$\iota^{\text{upper}}$  \\
		\midrule[0.8pt]
		\texttt{IMRPhenomXP} &  0.49 &14\% & 10\% &40\degree \\
		\texttt{IMRPhenomXHM} & 0.47 & 10\% & 5\% & 32\degree \\
		\bottomrule[1.1pt]
		\bottomrule[1.1pt]
	\end{tabular}
	}
    \caption{A summary of the results obtained simulating the parameter estimation of a non-spinning face-on BBH  with component masses of $29\, M_{\odot}$ and $36 \, M_{\odot}$, detected by ET at $\hat{\rho}=$100. The injections have been made via \texttt{IMRPhenomXHM}, the recoveries via both \texttt{IMRPhenomXHM} and \texttt{IMRPhenomXP}. The 95\% credible upper limit on the inclination angle is taken.}
	\label{tab_ET_q08}
\end{table}

\begin{table}[t] 
	\centering
	\scalebox{0.9}{
	\begin{tabular}{@{\hspace{0.2cm}}c@{\hspace{0.4cm}}c@{\hspace{0.4cm}}c@{\hspace{0.4cm}}c@{\hspace{0.4cm}}c@{\hspace{0.2cm}}}
	        \addlinespace[0.5em]
	        \toprule[1.1pt]
	        \toprule[1.1pt]
		Recovery model &  $\delta_{\text{st}}(d_L)$ & $\Delta d_L/d_L$ & $|d_L^{\text{inj}}-d_L^{\text{med}}|/d_L^{\text{inj}}$ &$\iota^{\text{upper}}$  \\
		\midrule[0.8pt]
		\texttt{IMRPhenomXP} &  0.49 &14\% & 10\% &40\degree \\
		\texttt{IMRPhenomXHM} & 0.48 & 14\% & 7\% & 37\degree \\
		\bottomrule[1.1pt]
		\bottomrule[1.1pt]
	\end{tabular}
	}
    \caption{A summary of the results obtained simulating the parameter estimation of a non-spinning face-on BNS  with component masses of $1.4\, M_{\odot}$ and $1.1 \, M_{\odot}$, detected by ET at $\hat{\rho}=$100, without considering tidal effects. The injection is performed with \texttt{IMRPhenomXHM}, while the recoveries use \texttt{IMRPhenomXHM} and \texttt{IMRPhenomXP}. The 95\% credible upper limit on the inclination angle is taken.}
	\label{tab_ET_BNS_q08}
\end{table}

\begin{figure} [t] 
	\centering
	\includegraphics[width=\columnwidth]{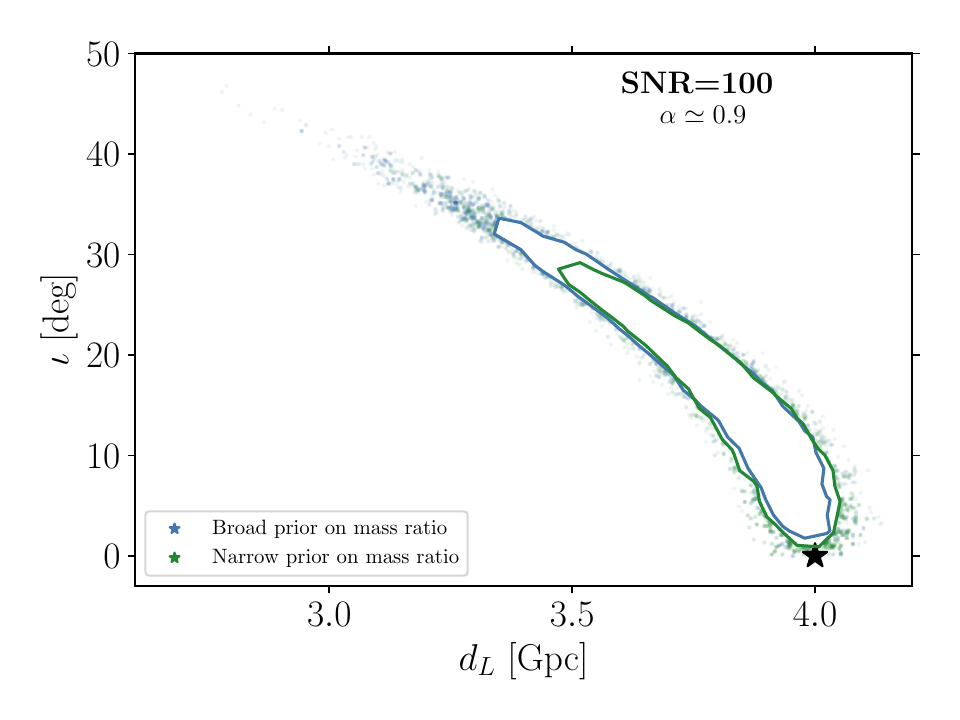}
	\caption{90\% contour plots for a face-on binary with masses $29\, M_{\odot}$ and $36 \, M_{\odot}$, detected by ET at an SNR of 100. The signals are injected and recovered with \texttt{IMRPhenomXHM}. The blue and green contours refer to simulations with a wide and a narrow prior on the mass ratio, respectively.}
	\label{ET_effect_massratio}  
\end{figure}  

\subsubsection{Next-to-unity mass ratio binaries II: the effects of biased estimation of the mass ratio}

We have highlighted that, as the mass ratio deviates further from unity, the higher-order modes become stronger, and the polarization amplitudes become more distinguishable (cfr.~Fig.\,\ref{fig_distinguishability}). Consequently, this effect could break the degeneracy between distance and inclination~\cite{Mills:2020thr}. On the other hand, ignoring those effects in the signal can lead to biases in terms of parameter estimation (see~\cite{Yi:2025pxe} in the context of LISA and~\cite{Gupta:2024gun} in the context of confusion with possible deviations from General Relativity). The results for the BBH (and for the two BNS) injections with $q_{\rm inj}=0.8$ using the \texttt{IMRPhenomXHM} approximant for both injection and recovery indicate that the mass ratio measurement appears to be biased. The posterior distributions suggest that the mass ratio is poorly recovered, tending towards values close to 1 as if the analysis were favouring the suppression of effects from higher-order multipoles. For instance, the complete inference of the BBH system simulated at $\hat{\rho}=$100, the posteriors of which for distance and inclination are shown in Fig.\,\ref{fig_ET_upperVSsnr_XHM_XHM_q08}, leads to a mass ratio of $q=0.90^{+0.09}_{-0.13}$, written as median value with 90\% credible interval bounds; note that half the samples are at $q>0.9$, where high-order modes are even weaker. This bias has three main negative consequences.
\begin{enumerate}
\item Measuring a next-to-unity mass ratio means suppressing higher-order modes, and this causes an additional amplification of the degeneracy between distance and inclination, as long as the system is optimally oriented ($\iota=0\degree$), or, combinations of the SNR and the alignment factor impact the recoveries negatively.
\item Regardless of the issue of the distance--inclination degeneracy, this bias would make the recovery of the component masses more difficult, directly bringing uncertainties to the source-frame masses, and indirectly bringing them to the redshift, via the degenerate luminosity distance if no electromagnetic information is available.
\item It would suppress effects that are enhanched in low-mass-ratio systems. Besides high-order modes, one can use precession as another channel to assist in breaking the degeneracy~\cite{Crescimbeni_inprep}.
\end{enumerate}

To investigate this aspect, we repeated the same face-on $36 \, M_{\odot}$--$29\, M_{\odot}$ injection at $\hat{\rho}=100$, but using a narrower prior on the mass ratio.
In Fig.\,\ref{ET_effect_massratio} we compare the 90\% $d_L$--$\iota$ credible regions for the two configurations, with both injection and recovery performed with \texttt{IMRPhenomXHM}. In Table~\ref{tab_ET_effect_massratio} we instead summarize the main results, which witness slight improvements when narrowing the prior on mass ratio. We add that the bias worsens as the SNR increases: for $\hat{\rho}=20$, the parameter estimation provides a mass ratio $q=0.83^{+0.15}_{-0.21}$. The increased SNR thus eliminates mass ratio samples closer to the injected value.

\begin{figure*} [t!] 
\centering
\includegraphics[width=\columnwidth]{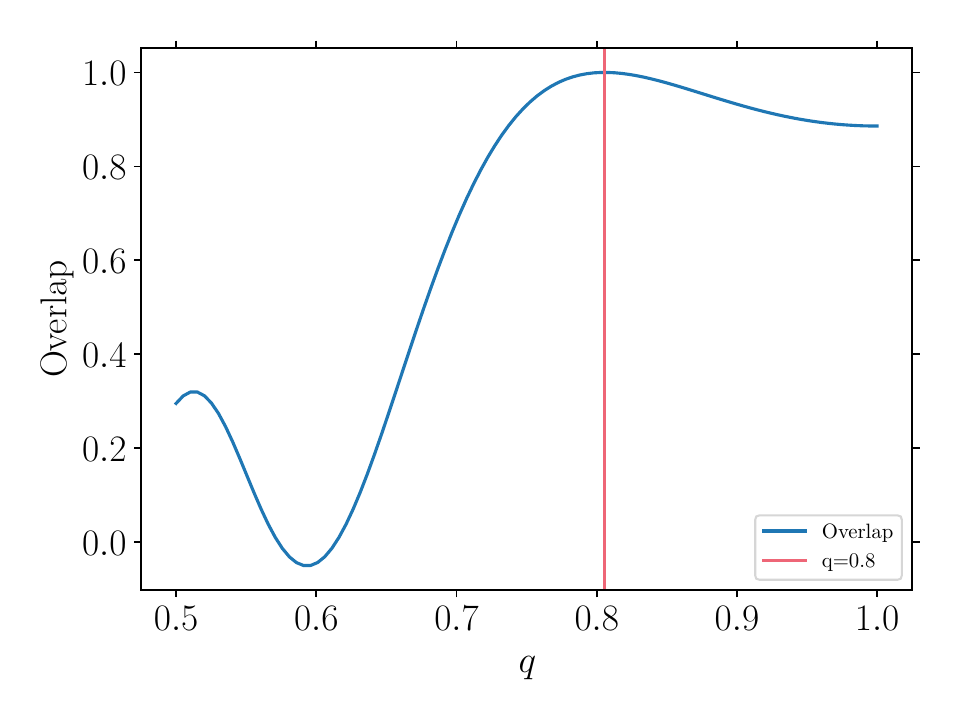}~~
\includegraphics[width=\columnwidth]{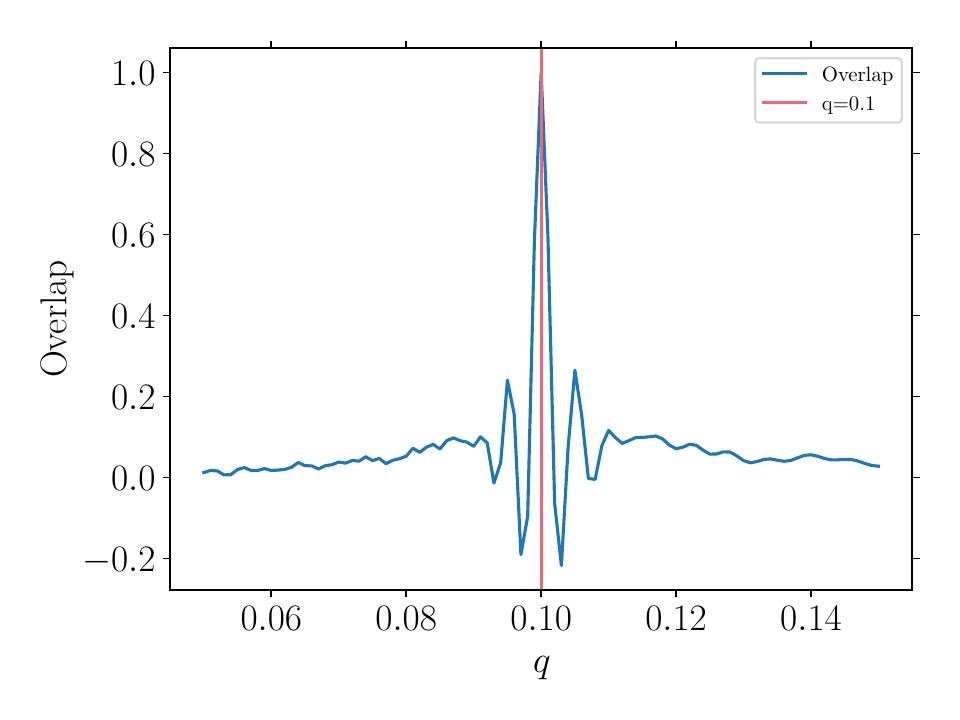}
\caption{Left panel: the  \textit{overlap} between a fixed linearly polarized waveform with mass ratio $q\sim 0.8$, indicated by the red vertical line, and waveforms with the same parameters but with varying mass ratio, according to Eq.\,(\ref{match_q}). The chirp mass is fixed at $\mathcal{M}=28.1 \, M_{\odot}$. Right panel: similar to the left panle, but the fixed mass ratio is $q=0.1$, while the chirp mass is fixed at $\mathcal{M}=14.8 \, M_{\odot}$ and the other parameters follow the left table of Table~\ref{tab_parameters_1}. All waveforms are generated with \texttt{IMRPhenomXHM}.}
\label{match_bias_XHM_q08}  
\end{figure*}

To understand the causes of this behaviour, we evaluate the \textit{overlap}, defined in Eq.\,(\ref{match}), between a waveform with a fix set of parameters $\bm{\mu}$ including the the mass ratio $q_{\text{inj}}$ and one with the same values of $\bm{\mu}$ but a varying mass ratio $q$.  This is motivated by the fact that the overlap enters the likelihood.  Formally, we calculate the following function of mass ratio:  
\begin{equation}\label{match_q}\begin{split}
		O(q) & = \, O[h(q,\bm{\mu}),h(q_{\text{inj}}, \bm{\mu})]= \\
		= \underset{\{\Phi_0, t_c\}}{\max} \, & \frac{ \, \left( \, h(q,\bm{\mu}) \, | \, h(q_{\text{inj}}, \bm{\mu}) \, \right)}{\sqrt{ \, \left( \, h(q,\bm{\mu})| \, h(q,\bm{\mu}) \, \right) \, \left( \, h(q_{\text{inj}}, \bm{\mu}) \, | \, h(q_{\text{inj}}, \bm{\mu}) \, \right) }} \,.
\end{split}\end{equation}
We first consider a ``fixed'' signal with again $36\,M_{\odot}$ and $29\,M_{\odot}$ for the detector frame component masses, that is, a fixed mass ratio of $q_{\text{inj}} \simeq 0.8$, $\iota=0\degree$, and vanishing spins.  The chirp mass of the varying waveforms is set to the one of the fixed signal, that is $\mathcal{M}=28.1 \, M_{\odot}$: indeed, we expect almost every sample to have a chirp mass close to the true one, since the chirp mass is measured with a precision of more than 1\textperthousand~in this configuration. In the left panel of Fig.\,\ref{match_bias_XHM_q08} we plot the function $O(q)$ for this specific comparison.   
When the high-order modes are included, starting from the injected value, where $O(q\simeq0.8)=1$, and going rightwards to $q=1$, the overlap remains almost constant, with $O(q=1)\sim 0.9$, while, when going leftwards, the function goes more rapidly to zero, and already at $q=0.7$ we have $O(q=0.7)
\sim 0.6$. The credible interval of the 90\% mass ratio is basically this region of almost constant $O(q)$, with next-to-unity values that are more probable, given the asymmetry of the credible interval itself. This might happen because, when the mass ratio approaches unity, the high-order terms are almost suppressed. This in turn interplays with the luminosity distance-inclination degeneracy, which enlarges the parameter space compatible with this configuration. On the other hand, the low-mass-ratio configuration is not biased. The complete parameter estimation for a face-on, $60 \, M_{\odot}$--$6 \, M_{\odot}$ system detected with an optimal SNR of 40 (i.e., the green distribution in Fig.\,\ref{fig_contours_ET_q01}) leads to $q=0.10^{+0.02}_{-0.01}$. Indeed, the \textit{overlap} of Eq.\,(\ref{match_q}) is sharply peaked around the $q=q_{\text{inj}}=0.1$ and drops rapidly and symmetrically to zero, as shown in the right panel of Fig.\,\ref{match_bias_XHM_q08}.

\begin{table}[t] 
	\centering
	\scalebox{0.9}{
	\begin{tabular}{@{\hspace{0.2cm}}c@{\hspace{0.4cm}}c@{\hspace{0.4cm}}c@{\hspace{0.4cm}}c@{\hspace{0.4cm}}c@{\hspace{0.2cm}}}
	        \addlinespace[0.5em]
	        \toprule[1.1pt]
	        \toprule[1.1pt]
		Prior on $q$ &  Recovered $q$ & $\Delta d_L/d_L$ & $|d_L^{\text{inj}}-d_L^{\text{med}}|/d_L^{\text{inj}}$ &$\iota^{\text{upper}}$  \\
		\midrule[0.8pt]
		Broad &  $0.90^{+0.09}_{-0.13}$ & 10\% & 5\% &32\degree \\
		Narrow & $0.81^{+0.03}_{-0.05}$ & 8\% & 3\% & 27\degree \\
		\bottomrule[1.1pt]
		\bottomrule[1.1pt]
	\end{tabular}
	}
    \caption{A summary of the results obtained simulating the parameter estimation of a non-spinning face-on BBH with component masses of $36\, M_{\odot}$ and $29 \, M_{\odot}$, detected by ET at $\hat{\rho}=$100. The same injection is recovered with a uniform broad prior in the range $[0.3,1]$ and a narrower prior in the range $[0.75,0.85]$; in both cases the injection and recovery is performed with \texttt{IMRPhenomXHM}. The 95\% credible upper limit on the inclination angle is taken.}
	\label{tab_ET_effect_massratio}
\end{table}

\section{Discussion and conclusions}

This work has rigorously quantified the challenges inherent in simultaneously inferring the luminosity distance and inclination of compact binaries in nearly face-on or face-off configurations, even with increased optimal SNR and network alignment factor --~conditions that yield robust constraints for more inclined systems.  We have systematically explored the role of GW higher-order modes in mitigating this degeneracy, revealing compelling results.  Specifically, we demonstrated that 3G detectors, with their enhanced sensitivity, will dramatically improve parameter estimation for asymmetric mass ratio binaries through the detection of higher-order modes.  While the improvement is less pronounced for binaries with mass ratios closer to unity, a measurable enhancement is still achieved.

Building upon this detailed study of higher-order modes, future research will investigate the impact of precession~\cite{Crescimbeni_inprep}. This analysis presents its own set of challenges, primarily due to the longer timescale of precession relative to orbital variations~\cite{Gerosa:2015tea}, requiring observations spanning many orbital cycles. 3G detectors will be crucial in this endeavor, enabling the observation of signals from lower frequencies and providing access to longer data stretches.  Similar to our findings with higher-order modes, the effects of precession are expected to be more prominent in nearly edge-on binaries~\cite{Vitale:2014mka} and low-mass-ratio systems.  Optimal combinations of SNR, spin, and mass ratio are anticipated to yield sufficient effectiveness in the parameter estimation.

Finally, we emphasize that this work mostly assumed ET in its standard triangular configuration, finding preliminary evidence that the degeneracy will be reduced by a 2L design, mostly due to the larger SNR. Future investigations should explore how alternative ET designs~\cite{Branchesi:2023mws, Maggiore:2024cwf} might influence the systematic effects detailed herein. Regardless of the specific ET design, the results will largely depend on the nature of the source. While BBHs with high-order modes can, in principle, help break the degeneracy, omitting these modes will reintroduce it. This will make the case of BNSs without an electromagnetic counterpart particularly challenging, as higher-order modes and precession are generally negligible.

\begin{acknowledgments}
We thank Alessandro Agapito for useful comments on the draft. We also thank Osvaldo Freitas, Francesco Iacovelli, Michele Maggiore, Niccolò Muttoni, and Michael Williams for internal reviewing and checking the draft. F.C. acknowledges the financial support provided under the ``Progetti per Avvio alla Ricerca Tipo 1'', protocol number AR12419073C0A82B. C.P. is supported by  ERC Starting Grant No.~945155--GWmining,  Cariplo Foundation Grant No.~2021-0555,  MUR PRIN Grant No.~2022-Z9X4XS,  MUR Grant ``Progetto Dipartimenti di Eccellenza 2023-2027'' (BiCoQ), and the ICSC National Research Centre funded by NextGenerationEU. P.P. acknowledges the financial support provided under the European Union's H2020 ERC, Starting Grant agreement no.~DarkGRA--757480. F.P.~acknowledges support from the ICSC - Centro Nazionale di Ricerca in High-Performance Computing, Big Data and Quantum Computing, funded by the European Union - NextGenerationEU and support from the Italian Ministry of University and Research (MUR) Progetti di ricerca di Rilevante Interesse Nazionale (PRIN) Bando 2022 - grant 20228TLHPE - CUP I53D23000630006. We also acknowledge support under the MIUR PRIN and FARE programmes (GW-NEXT, CUP:~B84I20000100001), and from the Amaldi Research Center funded by the MIUR program ``Dipartimento di Eccellenza'' (CUP: B81I18001170001), and by the INFN TEONGRAV initiative. Some numerical computations were performed at the Vera cluster supported by MUR and Sapienza University of Rome.
\end{acknowledgments}



\newpage 
\bibliography{Bibliography}
\end{document}